\documentclass[twocolumn,english,pra, showpacs]{revtex4}
\usepackage[T1]{fontenc}
\usepackage[latin9]{inputenc}
\usepackage{amsmath}
\usepackage{graphicx}
\usepackage{amssymb}

\makeatletter

\providecommand{\tabularnewline}{\\}

\@ifundefined{textcolor}{}
{%
 \definecolor{BLACK}{gray}{0}
 \definecolor{WHITE}{gray}{1}
 \definecolor{RED}{rgb}{1,0,0}
 \definecolor{GREEN}{rgb}{0,1,0}
 \definecolor{BLUE}{rgb}{0,0,1}
 \definecolor{CYAN}{cmyk}{1,0,0,0}
 \definecolor{MAGENTA}{cmyk}{0,1,0,0}
 \definecolor{YELLOW}{cmyk}{0,0,1,0}
 }

\makeatother

\usepackage{babel}

\begin{document}

\title{Zero-temperature phase diagram of Bose-Fermi gaseous mixtures in
optical lattices}

\author{T. P. Polak}

\address{Faculty of Physics, Adam Mickiewicz University of Pozna\'{n}, Umultowska
85, 61-614 Pozna\'{n}, Poland}

\email{tppolak@amu.edu.pl}

\author{T. K. Kope\'c}

\address{Institute for Low Temperatures and Structure Research, Polish Academy
of Sciences, P.O. Box 1410, 50-950 Wroclaw 2, Poland}
\begin{abstract}
We study the ground state phase diagram of a mixture of bosonic and
fermionic cold atoms confined on two- and three-dimensional optical
lattices. The coupling between bosonic fluctuations and fermionic
atoms can be attractive or repulsive and has similarities with electron-phonon
coupling in crystals. We investigate behavior of the mixtures in the
limit, where the Bogoliubov sound velocity that dictates bosonic dynamics
is comparable to the Fermi velocity, hence the retardation effects
are important part of the physics. The dynamic Lindhard response function
of the fermionic density to changes in the bosonic number of particles
above some critical frequency can alter the sign and in consequence
the inter-species interaction between particles becomes repulsive
in contrast to the static limit (instantaneous and always attractive).
Considering the above we show that the structure of the phase diagrams
crucially depends on the difference in masses of the bosons and fermions.
We discuss the situations where integrating out fermionic field provides
an additional interaction that can decrease or increase bosonic coherence.
\end{abstract}


\maketitle

\section{Introduction}

Trapping and cooling Bose-Fermi mixtures of dilute quantum gases has
opened a wide area of research in atomic physics. The interactions
between bosonic and fermionic species interconnect two systems of
fundamentally different quantum statistics. The diluteness of the
gaseous mixtures allows one to treat the interactions between particles
in terms of binary collisions. In consequence we can replace the real
inter-atomic potential by a pseudo-potential characterized by only
one parameter, the $s$-scattering length. The latter is experimentally\cite{ferlaino,gunter,ospelkaus,best}
tunable by exploiting optically or magnetically induced Feshbach resonances\cite{inouye}.
Despite its simplicity the interaction potential (mathematically ill
defined\cite{galindo}) of the ultra-cold multi-component gases confined
in optical lattices is responsible for a wealth of novel quantum phases\cite{lewenstein,albus}
including charge density waves (CDW)\cite{mering,mathey}, as well
as supersolid behavior\cite{buchler,titvinidze,orth}. The nature
of the phase transition and qualitative phase diagram for one-component
bosonic system can be inferred based on very simple arguments\cite{fisher}.
When tunneling between lattice sites of the bosons is suppressed,
compared with point-like interaction between them, the system can
undergo a quantum phase transition between a superfluid (SF) phase
(characterized by large number fluctuations at each lattice site),
and a Mott insulating (MI) phase where each lattice site is occupied
by precisely an integer number of bosons without any number fluctuations.
Adding to a bosonic system a fermionic ingredient and allowing for
the mutual repulsion or attraction between species of different statistics
strongly affects the equilibrium properties. Increasing the boson-fermion
repulsion drives the system towards spatial separation whereas attraction
gives rise to implosion\cite{capuzzi,akdeniz}. The dynamics underlying
the phase transitions in the Bose-Fermi mixtures is produced by the
small changes of the bosonic density which induce a modulation of
the fermionic density. As a consequence of the feedback of the fermionic
perturbation a shift of the bosonic energy occurs, thereby inducing
an additional attraction or repulsion that changes the original interaction
between bosons\cite{mazzarella}. In the case of deep optical lattices
and small densities, the coherent description of the system provided
by the Gross-Pitaevskii equation is not reliable due to rising effects
of correlation. The experimental data clearly demonstrated that adding
a fermionic cloud to strongly interacting bosons always results in
a decay of visibility of the interference pattern in time-of-flight
images\cite{best,gunter,ospelkaus}. Moreover, the scale of disappearance
of the coherence in the mixtures is different for attractive and repulsive
scattering lengths\cite{best}. To predict such behavior theoretically
one can include the more general, than one-component Bose-Hubbard
(BH), multi-band model. In mentioned approach if the higher-band renormalization
of the boson parameters is dominant over the fermion screening of
the interaction, the Mott-insulating lobes in the Bose-Hubbard phase
diagram are enhanced for either sign of the Bose-Fermi interactions\cite{lutchyn,tewari}.
On the other hand inclusions of the retardation effects\cite{akdeniz-1}
(which arises from the presence of very low energy excitations in
a Fermi sea) give rise to so-called orthogonality catastrophe\cite{refael}.
Another approach to quantum mixtures of particles of unequal masses,
when the difference in the tunneling amplitudes between heavy bosons
and light fermions is large enough to neglect quantum nature of the
bosons, provides to description of the system in Fermi-Bose version
of the Falicov-Kimball model\cite{iskin}. The mutual interactions
of bosons and fermions can affect the spectrum of collective excitations
in the collisionless regime as the mixture goes toward either demixing
or collapse\cite{capuzzi}. It has been shown that mode-mode coupling
effects may arise when sound velocity of the Bose gas is comparable
to the Fermi velocity of the fermions\cite{yip}. The energy spectra
of the bosonic and fermionic mixtures and phase diagram were also
obtained by the field theory methods\cite{han}, however calculated
phase boundary does not change the structure and only shifts the chemical
potential. The properly constructed effective theory lead to an effective,
fermion mediated, long-range interaction between bosons with alternating
sign that is the origin of the CDW and can explain the MI-CDW phase
separation\cite{mering}. There is also another possibility of the
analysis of the mixtures of atoms with different statistics where
the second species is strongly localized on random sites which can
lead to random shifts of the on-site energies and, in consequence,
the disorder with discrete probability distribution is created \cite{krutitsky}.
Recently the experiments \cite{catani} on a harmonically trapped
mixtures of atomic bose-bose gases show that the presence of relevant
fraction of the $^{41}$K bosonic species modifies the quantum phase
transition occurring in \emph{$ $}$^{87}$Rb inducing a significant
loss of coherence similarly to bose-fermi systems that can be explained
in the mean-field theory framework\cite{buosante}.

The aim of this work is to study the superfluid to Mott-insulator
zero-temperature phase transition by means of the Bose-Fermi-Hubbard
model in two- ($2D$) and three-dimensional ($3D$) optical condensates.
In order to find a phase boundary for BF mixtures very sophisticated
methods and calculations are required. Only several theoretical works
concentrated on the inherent difficulty of dealing with BF Hubbard
Hamiltonian originates from the non-perturbative nature of the model
and retardation effects. To elucidate the quantum phase transition
in optical lattices, where the kinetic energy scale is less than the
dominating repulsive energy and density-density coupling between species
with different statistics comes into play, we have adopted a theoretical
approach for strongly interacting fermions \cite{kopec} to the BF
Hubbard model in a way to include the effects of particle number fluctuations
and make the qualitative phase diagrams more quantitative \cite{polak}.
To facilitate this task, we employ a functional integral formulation
of the theory that enables to perform functional integration over
fields defined on different topologically equivalent classes of the
$\mathrm{U}\left(1\right)$ group, i.e., with different winding numbers.
An inclusion of the winding numbers (comes from periodicity of the
phase variables) is unavoidable in order to properly construct the
phase diagram and the Poisson re-summation formula turns out to be
very useful for derivation of the topological term of the partition
function. The quantum rotor representation method we use is deeply
rooted in the gauge symmetries of the model. We construct an invariant
theory introducing an appropriate $\mathrm{U}\left(1\right)$ gauge
transformation. In Sec. II we review the Hamiltonian for the system
and show the connections of the parameters to the experimentally measured
quantities in optical lattices. Sec. III contains description of the
method we use and can serve a guidance to obtain the critical line
equation presented in Sec. IV. Before showing the phase diagrams for
the Bose-Fermi Hubbard model in the quantum rotor description we make
some general remarks in Sec. V concern the phase boundary equation
and compare our results with the diagrammatic perturbation approach
to the one-component Bose-Hubbard model for experimentally accessible
densities of the particles. Sec. VI presents the discussion of the
ground state phase diagrams for the Bose-Fermi Hubbard Hamiltonian
calculated within quantum rotor approach. The Sec. VII. is devoted
to some concluding remarks. The appendixes contain the derivation
of relevant formulas of the main text and are introduced to keep the
text self-contained.

\section{Hamiltonian}

For bosons confined in optical lattices the two main energy scales
are set by the hopping amplitude proportional to $t_{b}$ (which sets
the kinetic energy scale for bosons) due to the particles tunneling,
and the on-site interaction $U_{b}>0$. For $t_{b}>U_{b}$ the phases
of the superfluid order parameter on individual lattice sites are
well defined. On the other hand, for sufficiently large repulsive
energy $U_{b}$, the quantum phase fluctuations lead to complete suppression
of the long-range phase coherence even at zero temperature. The competition
between the kinetic energy, which is gained by delocalizing bosons
over lattice sites and the repulsive interaction energy, which disfavors
having more than one particle at any given site, can be modeled by
quantum Bose-Hubbard Hamiltonian\cite{fisher}. The physics of the
bosonic and non-interacting spin-polarized (collisions in the $s$-wave
channel are forbidden by their statistics) fermionic mixtures with
density-density interaction between species $U_{bf}$ leads to Bose-Fermi-Hubbard
Hamiltonian \cite{albus}: \begin{eqnarray}
\mathcal{H} & = & \frac{U_{b}}{2}\sum_{i}n_{bi}^{2}-\sum_{\left\langle i,j\right\rangle }t_{bij}b_{i}^{\dagger}b_{j}-\bar{\mu}_{b}\sum_{i}n_{bi}\nonumber \\
 & - & \sum_{\left\langle i,j\right\rangle }t_{fij}f_{i}^{\dagger}f_{j}-\mu_{f}\sum_{i}n_{fi}+U_{bf}\sum_{i}n_{bi}n_{fi},\label{hamiltonian}\end{eqnarray}
where $b_{i}^{\dagger}\left(f_{i}^{\dagger}\right)$ and $b_{i}\left(f_{i}\right)$
stand for the bosonic (fermionic) creation and annihilation operators
$n_{bi}=b_{i}^{\dagger}b_{i}$, $\left(n_{fi}=f_{i}^{\dagger}f_{i}\right)$
is the boson (fermion) number operator on the site $i$, and the reduced
chemical potential $\bar{\mu}_{b}=\mu_{b}+U_{b}/2$ controls the number
of bosons and $\mu_{f}$ fermions respectively. Here, $\left\langle i,j\right\rangle $
identifies summation over the nearest-neighbor sites. Furthermore,
$t_{bij}\left(t_{fij}\right)$ is the hopping matrix element for bosons
(fermions). For simplicity, we neglect the inhomogeneous magnetic
trap potential. If the on-site boson-fermion coupling strength $U_{bf}$
becomes very strong the dilute gaseous mixtures are unstable to phase
separation $\left(U_{bf}>0\right)$ or to collapse of the phase separated
configuration $\left(U_{bf}<0\right)$ \cite{viverit,gunter}. The
presence of the lattice will introduce kinetic energy scales $t_{b\left(f\right)}$
competing with $U_{bf}$ stabilizing the system. We assume that an
optical lattice created by the counter-propagating laser beams is
deep enough and we can restrict ourselves to the lowest Bloch bands.
The corresponding experimental parameters can be estimated by following
relations\cite{titvinidze}\begin{eqnarray}
t_{x} & \simeq & \frac{4}{\sqrt{\pi}}E_{r}^{x}\left(\frac{V_{0}}{E_{r}^{x}}\right)^{3/4}\exp\left[-2\left(\frac{V_{0}}{E_{r}^{x}}\right)\right],\\
U_{x} & \simeq & \sqrt{\frac{8}{\pi}}ka_{x}E_{r}^{x}\left(\frac{V_{0}}{E_{r}^{x}}\right)^{3/4},\end{eqnarray}
(subscript $x=\left\{ b,f\right\} $ means $b$ bosons and $f$ fermions
respectively) where boson-boson $a_{b}$, fermion-fermion $a_{f}$
and boson-fermion $a_{bf}$ \begin{equation}
U_{bf}\simeq\sqrt{\frac{8}{\pi}}ka_{bf}E_{r}^{b}\left(\frac{V_{0}}{E_{r}^{b}}\right)^{3/4}\frac{1+\frac{m_{b}}{m_{f}}}{\left(1+\sqrt{\frac{m_{b}}{m_{f}}}\right)^{3/2}}\label{ubfexp}\end{equation}
scattering lengths can be continuously tune in the experiments\cite{ospelkaus,ferlaino,gunter,best}
inducing attractive or repulsive interaction between species. The
$k=2\pi/\lambda$ is the wavelength of the laser and $E_{r}^{x}=\hbar^{2}k^{2}/2m_{x}$
is the recoil energy and $m_{x}$ is the atomic mass.

\section{Description of the method}

We write the partition function of the system switching from the particle-number
representation to the conjugate phase representation of the bosonic
degrees of freedom using the bosonic and fermionic path-integral over
the complex fields $a_{i}\left(\tau\right)$ and $f_{i}\left(\tau\right)$
depending on the {}``imaginary time'' $0\leq\tau\leq\beta\equiv1/k_{\mathrm{B}}T$
with $T$ being the temperature:\begin{equation}
\mathcal{Z}=\int\left[\mathcal{D}\bar{b}\mathcal{D}b\mathcal{D}\bar{f}\mathcal{D}f\right]e^{-\mathcal{S}\left[\bar{b},b,\bar{f},f\right]}.\end{equation}
The action $\mathcal{S}$ is given by\begin{equation}
\mathcal{S}=\mathcal{S}_{B}\left[\bar{b},b,\bar{f},f\right]+\int_{0}^{\beta}d\tau\mathcal{H\left(\tau\right)},\end{equation}
where\begin{eqnarray}
\mathcal{S}_{B}\left[\bar{b},b,\bar{f},f\right] & = & \sum_{i}\int_{0}^{\beta}d\tau\bar{b}_{i}\left(\tau\right)\frac{\partial}{\partial\tau}b_{i}\left(\tau\right)\nonumber \\
 & + & \sum_{i}\int_{0}^{\beta}d\tau\bar{f}_{i}\left(\tau\right)\frac{\partial}{\partial\tau}f_{i}\left(\tau\right).\end{eqnarray}
In the next section we will integrate over the fermionic fields since
the action is quadratic in $f_{i}\left(\tau\right)$ variables. We
attempt to reduce the large number of degrees of freedom in the partition
function to the few which dominate the low energy physics.

\subsection{Integration over fermionic fields}

Before integrating out of the fermionic degrees of freedom we write
the action in the form:

\begin{eqnarray}
\mathcal{S}_{b}\left[\bar{b},b,n_{b}\right] & = & \int_{0}^{\beta}d\tau\left\{ \sum_{i}\left[\bar{b}_{i}\left(\tau\right)\frac{\partial}{\partial\tau}b_{i}\left(\tau\right)+\frac{U_{b}}{2}n_{bi}^{2}\left(\tau\right)\right]\right.\nonumber \\
 & - & \left.\sum_{\left\langle i,j\right\rangle }t_{bij}\bar{b}_{i}\left(\tau\right)b_{j}\left(\tau\right)-\bar{\mu}_{b}\sum_{i}n_{bi}\left(\tau\right)\right\} ,\nonumber \\
\mathcal{S}_{f}\left[\bar{f},f,n_{f}\right] & = & \int_{0}^{\beta}d\tau\left[\sum_{i}\bar{f}_{i}\left(\tau\right)\frac{\partial}{\partial\tau}f_{i}\left(\tau\right)\right.\nonumber \\
 & + & \left.\sum_{\left\langle i,j\right\rangle }t_{fij}\bar{f}_{i}\left(\tau\right)f_{j}\left(\tau\right)-\mu_{f}\sum_{i}n_{fi}\left(\tau\right)\right],\nonumber \\
\mathcal{S}_{\mathrm{int}}\left[n_{b},n_{f}\right] & = & U_{bf}\sum_{i}\int_{0}^{\beta}d\tau n_{bi}\left(\tau\right)n_{fi}\left(\tau\right).\end{eqnarray}
We notice that adding the inter-species interaction term to the fermionic
part of the action\begin{equation}
\mathcal{S}_{f+\mathrm{int}}=\sum_{i,j}\int_{0}^{\beta}d\tau d\tau'\bar{f}_{i}\left(\tau\right)\left[\hat{G}_{f+\mathrm{int}}\left(\tau,\tau'\right)\right]_{ij}f_{j}\left(\tau'\right).\end{equation}
allows one to integrate over fermionic fields, because the action
is Gaussian in $f_{i}\left(\tau\right)$ operators. Resulting partition
function takes the form\begin{equation}
\mathcal{Z}=\int\left[\mathcal{D}\bar{b}\mathcal{D}b\right]e^{-\mathcal{S}_{b}\left[\bar{b},b,n_{b}\right]}e^{\mathrm{Tr}\ln\hat{G}_{f+\mathrm{int}}^{-1}}.\label{partitionfunction1}\end{equation}
We will be looking for solutions obeying translational invariance
in the {}``imaginary time'' direction, i.e., such that the partition
function depends only on the difference $\left|\tau-\tau'\right|$.
Expanding the trace of the logarithm in Eq. (\ref{partitionfunction1})
we have \begin{eqnarray}
\mathrm{Tr}\ln\hat{G}_{f+\mathrm{int}}^{-1} & = & -\mathrm{Tr}\ln\hat{G}_{f}-\mathrm{Tr}\hat{G}_{\mathrm{int}}\hat{G}_{f}\nonumber \\
 & - & \frac{1}{2}\mathrm{Tr}\left(\hat{G}_{\mathrm{int}}\hat{G}_{f}\right)^{2},\end{eqnarray}
with\begin{eqnarray}
\left[\hat{G}_{f}^{-1}\left(\tau,\tau'\right)\right]_{ij} & = & \left[\left(\frac{\partial}{\partial\tau}-\mu_{f}\right)\delta_{ij}-t_{fij}I_{ij}\right]\delta\left(\tau-\tau'\right),\nonumber \\
\left[\hat{G}_{\mathrm{int}}\left(\tau,\tau'\right)\right]_{ij} & = & U_{bf}\bar{b}_{i}\left(\tau\right)b_{j}\left(\tau'\right)\delta_{ij}\delta\left(\tau-\tau'\right).\end{eqnarray}
We defined $I_{ij}=1$ if $i,j$ are the nearest neighbors and equals
zero otherwise. Trace over first term of the expansion gives constant
contribution of the fermions, in the non-interacting system, to the
action. Second one induces a shift in the chemical potential of bosons.
The third term after exploiting Fourier-Matsubara transform\begin{equation}
b_{i}\left(\tau\right)=\frac{1}{N\beta}\sum_{\boldsymbol{k},\ell}b_{\boldsymbol{k}}\left(\omega_{\ell}\right)e^{-i\left(\omega_{\ell}\tau-\boldsymbol{k}\cdot\boldsymbol{r}_{i}\right)},\end{equation}
where $\omega_{\ell}=2\pi\ell/\beta$ ($\nu_{\ell}=\pi\left(2\ell+1\right)/\beta$)
with ($\ell=0,\pm1,\pm2,...$) are Bose(Fermi)-Matsubara frequencies
respecting periodic (anti-periodic) boundary conditions of the bosonic
(fermionic) field operator, reduces to\begin{eqnarray}
\mathrm{Tr}\ln\hat{G}_{f+\mathrm{int}}^{-1} & = & \frac{U_{bf}^{2}}{2}\sum_{\boldsymbol{q},\ell}\Lambda_{\boldsymbol{q}}\left(\omega_{\ell}\right)\nonumber \\
 & \times & \chi\left(\boldsymbol{q},i\nu_{\ell},\mu_{f},\beta\right)\Lambda_{-\boldsymbol{q}}\left(-\omega_{\ell}\right),\label{trace}\end{eqnarray}
where $\Lambda_{\boldsymbol{q}}\left(\omega_{\ell}\right)=\bar{b}_{\boldsymbol{q}}\left(\omega_{\ell}\right)b_{\boldsymbol{q}}\left(\omega_{\ell}\right)$
and $\chi\left(\boldsymbol{q},i\nu_{\ell},\mu_{f},\beta\right)$ is
called Lindhard function%
\begin{figure}
\includegraphics[scale=0.9]{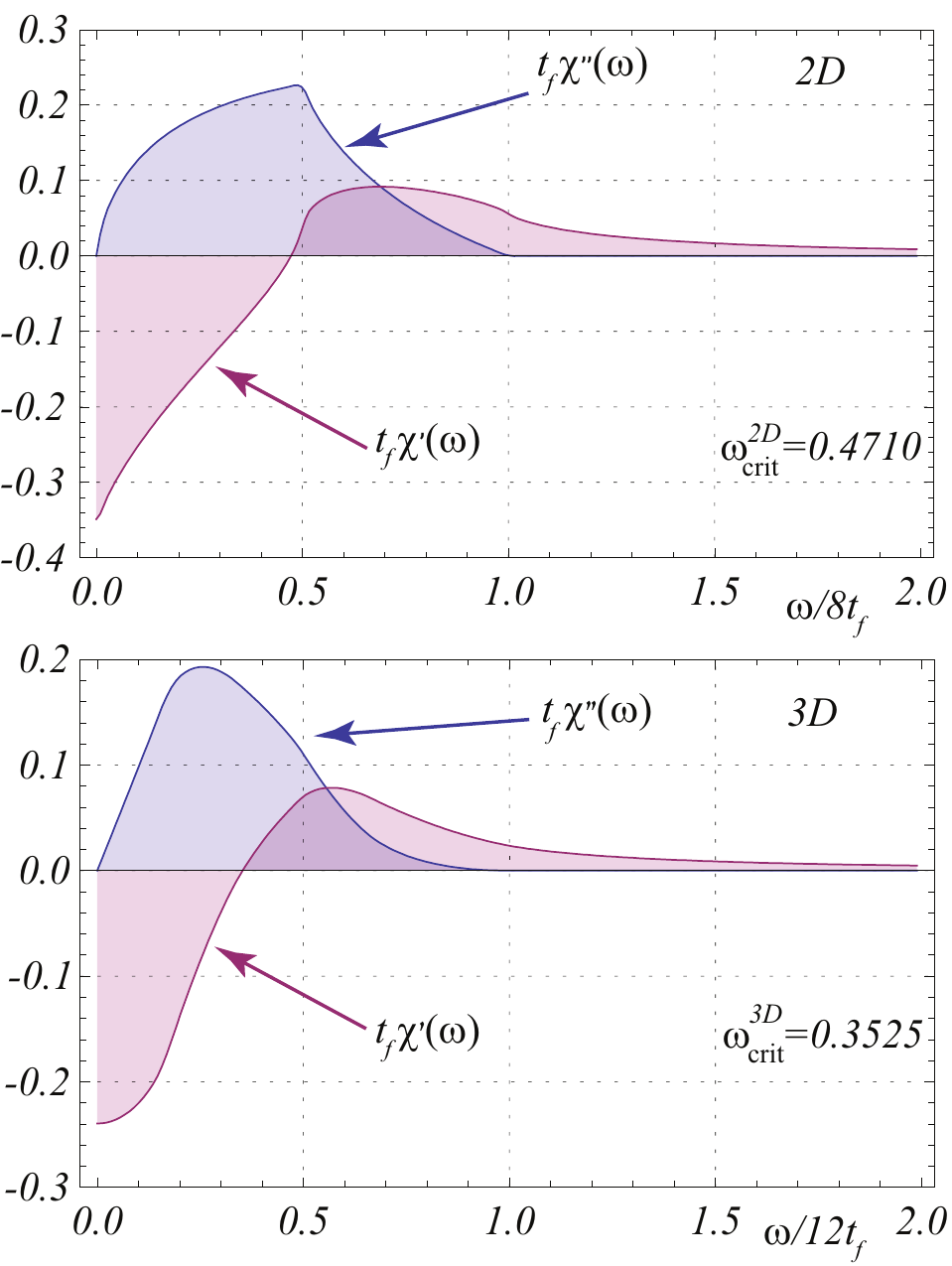}\caption{(Color online) Real $t_{f}\chi'\left(\omega\right)$ and imaginary
$t_{f}\chi''\left(\omega\right)$ part of the local dynamic Lindhard
function for square ($2D$) and cubic lattice ($3D$), in the low-temperature
limit, as a function of frequency, where the normalized fermionic
potential $\mu_{f}/t_{f}=0$ is equal zero. The normalized value of
the critical frequency $\omega_{\mathrm{crit}}^{2D,3D}$ shows where
induced, frequency-dependent, effective part of the interaction $U_{bf}^{2}\chi'\left(\omega_{\mathrm{crit}}\right)$
changes character from attractive to repulsive.}
\label{fig1}
\end{figure}
 \begin{equation}
\chi\left(\boldsymbol{q},i\nu_{\ell},\mu_{f},\beta\right)=\sum_{\mathbf{k}}\frac{f\left[t_{f\boldsymbol{k}},\mu_{f},\beta\right]-f\left[t_{f\boldsymbol{k}+\boldsymbol{q}},\mu_{f},\beta\right]}{t_{f\boldsymbol{k}}-t_{f\boldsymbol{k}+\boldsymbol{q}}-i\nu_{\ell}}.\label{lindhard}\end{equation}
In the above $f\left[t_{f\boldsymbol{k}},\mu_{f},\beta\right]=1/\left\{ \exp\left[\beta\left(t_{f\boldsymbol{k}}-\mu_{f}\right)\right]+1\right\} $
is the Fermi distribution function and $t_{f\boldsymbol{k}}$ is the
fermionic dispersion relation. To stay in the local (momentum integrated)
regime we perform $\boldsymbol{q}$ and \textbf{k} integration over
the first Brillouin zone and in the $T\rightarrow0$ limit using an
analytic continuation $i\nu_{\ell}\rightarrow$$\omega+i\epsilon$
we obtain imaginary part $\chi''\left(\omega,\mu_{f}\right)\equiv\mathrm{Im}\chi\left(\omega,\mu_{f}\right)$
of the local dynamic Lindhard function (see Appendix for details)\begin{eqnarray}
\chi''\left(\omega',\mu_{f}\right) & = & \int_{-\infty}^{+\infty}dx\left[\Theta\left(x-\omega'-\mu_{f}\right)-\Theta\left(x-\mu_{f}\right)\right]\nonumber \\
 & \times & \rho\left(x\right)\rho\left(x-\omega'\right)\end{eqnarray}
that satisfies sum rule which is just the conservation of the number
of particles. In the above $\rho\left(\xi\right)=N^{-1}\sum_{\mathbf{k}}\delta\left[\xi-t_{f\boldsymbol{k}}\right]$
is the density of states and $\Theta\left(x\right)$ is the unit step
function. Therefore, the corresponding real part $\chi'\left(\omega,\mu_{f}\right)\equiv\mathrm{Re}\chi\left(\omega,\mu_{f}\right)$
can be easily deduced from Kramers-Kr\"{o}nig relation \begin{equation}
\chi'\left(\omega,\mu_{f}\right)=\frac{2}{\pi}\int_{0}^{+\infty}\frac{\omega'\chi''\left(\omega',\mu_{f}\right)}{\omega'^{2}-\omega^{2}}d\omega'.\end{equation}
Later, without any loss of generality, we drop the chemical potential
dependence writing $\chi'\left(\omega,\mu_{f}=0\right)\equiv\chi'\left(\omega\right)$
and $\chi''\left(\omega,\mu_{f}=0\right)\equiv\chi''\left(\omega\right)$.
Finally the {}``imaginary-time'' partition function with integrated
out fermionic degrees of freedom, in the local approximation (see
Appendix for details), can be written as:\begin{equation}
\mathcal{Z}=\int\left[\mathcal{D}\bar{b}\mathcal{D}b\right]e^{-\mathcal{S}_{\mathrm{eff}}\left[\bar{b},b\right]}\end{equation}
with the effective action \begin{eqnarray}
\mathcal{S}_{\mathrm{eff}}\left[\bar{b},b\right] & = & \int_{0}^{\beta}d\tau\left\{ \sum_{i}\bar{b}_{i}\left(\tau\right)\frac{\partial}{\partial\tau}b_{i}\left(\tau\right)\right.\nonumber \\
 & + & \frac{U_{\mathrm{eff}}}{2}\sum_{i}n_{bi}^{2}\left(\tau\right)-\sum_{\left\langle i,j\right\rangle }t_{bij}\bar{b}_{i}\left(\tau\right)b_{j}\left(\tau\right)\nonumber \\
 & - & \left.\bar{\mu}_{b}\sum_{i}n_{bi}\left(\tau\right)\right\} \label{effective action}\end{eqnarray}
expressed in terms of bosonic degrees of freedom only. We want to
emphasize that applying a local, in the Matsubara-imaginary time,
approach we neglect any dissipation effects. Whereas locality in real
space rules out some parts of interesting physics such as the charge
density wave, namely an insulating phase with modulated density or
the supersolid phase, presenting the coexistence of superfluidity
and a periodic spatial modulation of the density, different from that
of the lattice. On the other hand, the long-range character of the
fermion mediated interaction between bosons with the fermion-induced
mean field potential can lead to spatially homogeneous regions of
commensurate CDW \cite{mering}. The motivation of the local approximation
was the idea that response of an interacting system can be pictured
as the response of a non-interacting system to an effective self-consistent
fi{}eld, that depends on global properties such as the particle densities.
A question of both fundamental and practical interest is, to what
extent can the physics of the exact non-local interaction be captured
by an approximate local theory? It seems that local approximations
often work surprisingly well, yielding energies very accurately, without
suffering from some of the characteristic drawbacks of non-locality
\cite{giuliani}.

From Eq. (\ref{effective action}) it is concluded that there is a
striking resemblance to the one-component Bose-Hubbard action with
the original repulsive interaction replaced now by \begin{equation}
U_{b}\rightarrow U_{\mathrm{eff}}=U_{b}+U_{bf}^{2}\chi'\left(\omega,\mu_{f}\right)\label{u effective}\end{equation}
which is the induced, frequency-dependent, effective interaction between
bosons. From Eq. (\ref{u effective}) we see that integrating out
fermionic field from BF Hubbard Hamiltonian provides an additional
interaction among bosons, which is \emph{not affected} by the attractive
or repulsive nature of the inter-species interaction $\pm U_{bf}$.
Before we proceed with further calculations let us make some remarks.
The substitution we introduced in Eq. (\ref{u effective}) is deceptively
simple and can lead to the assumption that the phase diagram of the
BFH model can be easily derived from the critical line of the one-component
BH Hamiltonian, which already has been obtained in several approximations.
Unfortunately, as we will see in the next section, $U_{bf}^{2}\chi'\left(\omega,\mu_{f}\right)$
is \emph{not} the only one ingredient to the final equation for the
critical line. The additional part, which influences the phase boundary
line condition, comes from the inter-species interaction and number
of fermions added to the system has an impact on amplitude of the
order parameter. Moreover, the chemical potential for fermions $\mu_{f}/t_{f}$
is shifted (we postpone calculations of it now and show proper formula
later) because of the induced effective interaction between them.
\begin{figure}
\includegraphics[scale=0.7]{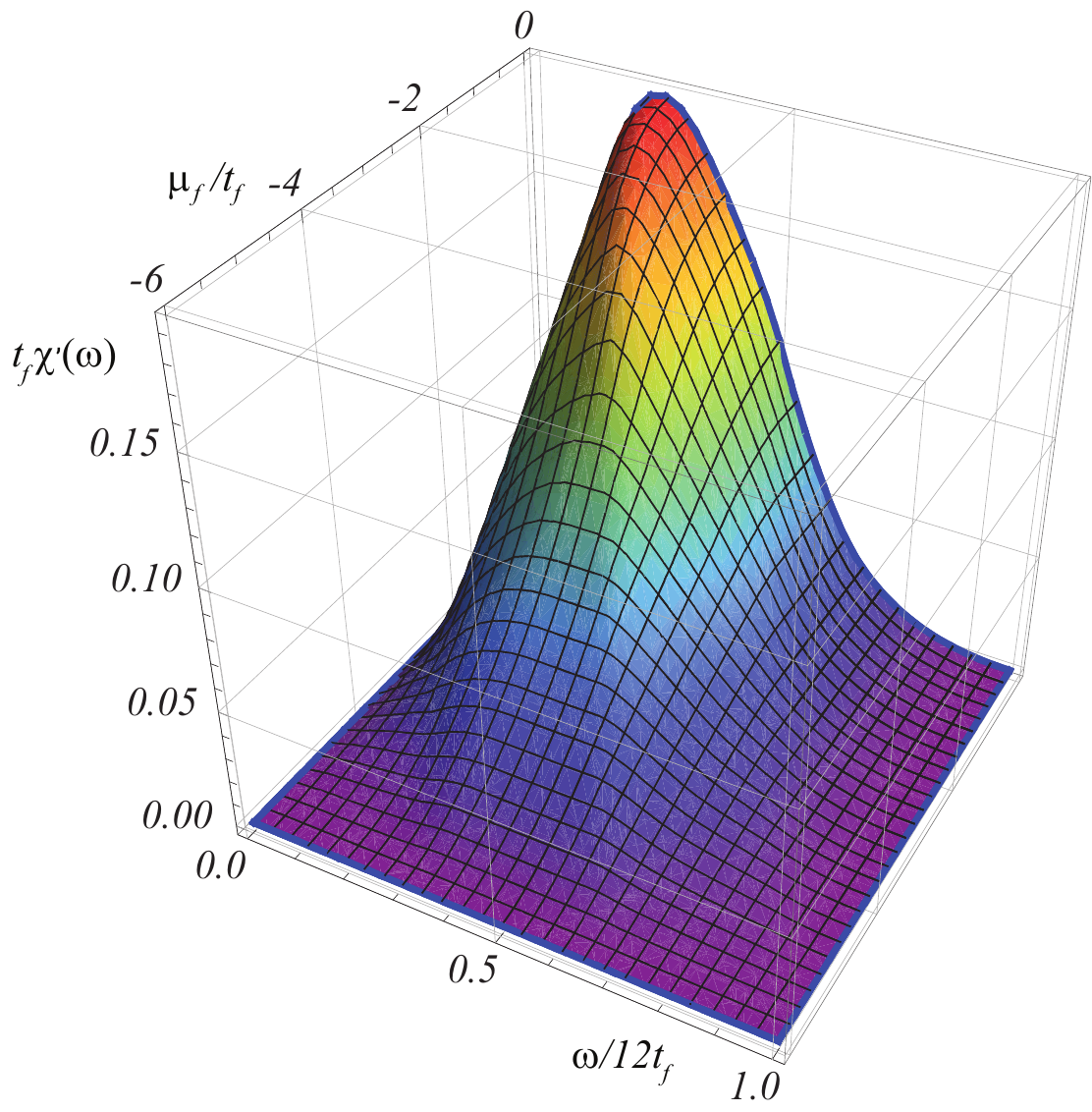}\caption{(Color online) Imaginary $t_{f}\chi''\left(\omega,\mu_{f}\right)$
part of the local dynamic Lindhard function for cubic lattice in the
space of the parameters: normalized frequency $\omega/t_{f}$ and
fermionic chemical potential $\mu_{f}/t_{f}$ in the zero-temperature
limit.}
\label{fig2}
\end{figure}
\begin{figure}
\includegraphics[scale=0.7]{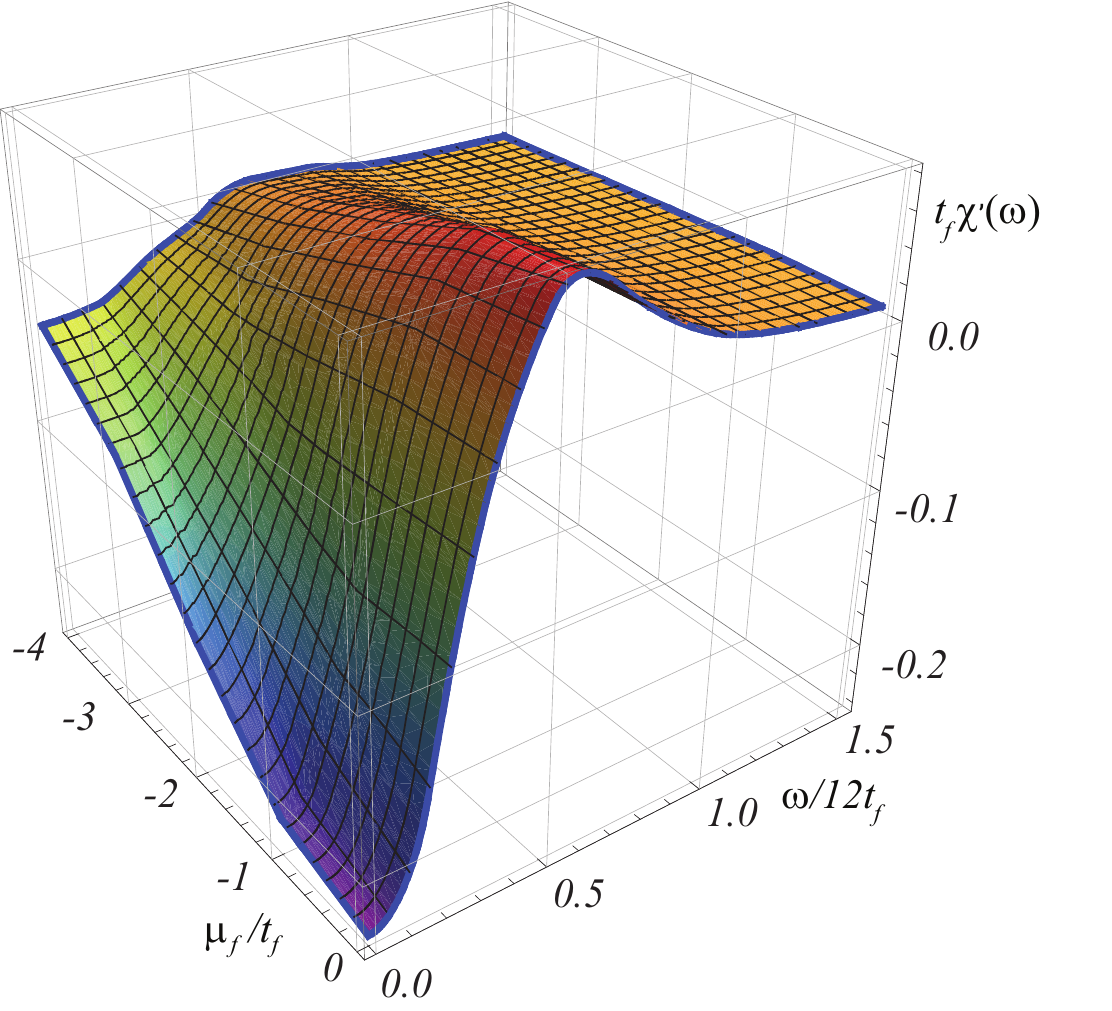}\caption{(Color online) Real $t_{f}\chi'\left(\omega,\mu_{f}\right)$ part
of the local dynamic Lindhard function for cubic lattice in the space
of the parameters: normalized frequency $\omega/t_{f}$ and fermionic
chemical potential $\mu_{f}/t_{f}$ in the zero-temperature limit.}
\label{fig3}
\end{figure}

Next step in the calculations depends on the ratio $m_{b}/m_{f}$
of the masses of bosons and fermions which can be seen from Eq. (\ref{ubfexp}).
Consequence of the latter is the fact that the speed of the Bogoliubov
sound $c{}_{b}$ for bosons differs from the first sound $v_{f}$
of the ideal Fermi gas. In typical experimental realizations $^{40}\mathrm{K}$-$^{87}\mathrm{Rb}$
systems the acoustic long-wavelength (boson) $c_{b}$ and fermion
$v_{f}$ velocities are comparable or $c_{b}/v_{f}<1$ (boson mass
is larger than fermion). Therefore we do not restrict our calculations
to the static limit but consider also local dynamical response function
(thus including the retardation effects). If $c_{b}$ is much larger
than $v_{f}$ (in the $^{40}\mathrm{K}$-$^{23}\mathrm{Na}$ species
we have $c_{b}/v_{f}\sim5$) the resulting interaction between bosons
is \emph{instantaneous} and \emph{always} attractive (with $U_{bf}^{2}\chi'\left(\omega,\mu_{f}\right)<0$)
so using the static approximation is justified with an error which
involves the small parameter $v_{f}/c_{b}$. 

The imaginary part of the Lindhard response function rises to a broad
peak before falling and the real part takes zero when changes in $\chi''\left(\omega\right)$
are the biggest (see Fig. \ref{fig1}). The real part $\chi'\left(\omega\right)$
is negative for the frequencies $\omega/8t_{f}<0.4710=\omega_{\mathrm{crit}}^{2D}$
for square and $\omega/12t_{f}<0.3525=\omega_{\mathrm{crit}}^{3D}$
for cubic lattice (see Fig. \ref{fig1}). We normalized the frequency
by the width of the band for non-interacting fermions to show the
scale of the energy. Nevertheless, for higher values of the normalized
frequency $\omega>\omega_{\mathrm{crit}}$ the induced part of the
interaction $U_{bf}^{2}\chi'\left(\omega\right)$ can be \emph{positive}
and \emph{increase} the effective interaction between bosons, in consequence
provide stronger localization these species on lattice sites. In the
case of large fermion hopping and commensurate filling with the lattice
the effective long-range density-density interaction between bosons
has alternating sign and is the origin of the charge density-wave
phases \cite{mering}. The higher values of the normalized chemical
potential for the fermions $\mu_{f}/t_{f}$ decreases the values of
both (Fig. \ref{fig2} and Fig. \ref{fig3}) real and imaginary part
of the local Lindhard function. That leads to the situation where
the terms containing explicitly the average density of fermions $n_{\mathrm{F}}$
will acquire more significance than terms with exclusively the inter-species
interaction $U_{bf}$.

\subsection{Static and periodic bosonic fields and gauge transformation}

Unfortunately the effective action is not quadratic in bosonic fields
$b_{i}$ and we have to decouple the effective interaction term in
Eq. (\ref{effective action}) by a Gaussian integration over the auxiliary
scalar potential fields \begin{equation}
V_{i}\left(\tau\right)=V_{i0}+V_{i}'\left(\tau\right),\end{equation}
with static \begin{equation}
V_{i0}=\frac{1}{\beta}V_{i}\left(\omega_{\ell=0}\right)\end{equation}
and periodic part \begin{equation}
V'_{i}\left(\tau\right)=\frac{1}{\beta}\sum_{\ell=1}^{+\infty}V_{i}\left(\omega_{\ell}\right)e^{i\omega_{\ell}\tau}+\mathrm{c.c},\label{periodic part V}\end{equation}
where $\omega_{\ell}$ is the Bose-Matsubara frequency. We observe
now that effective BF Hubbard Hamiltonian has a local $\mathrm{U}\left(1\right)$
gauge symmetry, when expressed in terms of the underlying boson variables.
This points out a possibility of an emergent dynamical $\mathrm{U}\left(1\right)$
gauge field as a fluctuating complex field attached to bosonic variables,
which is dynamically generated, by interacting bosons. Thus, the periodic
part $V'_{i}\left(\tau\right)\equiv V'_{i}\left(\tau+\beta\right)$
couples to the local particle number through the Josephson-like relation
$\dot{\phi}_{i}\left(\tau\right)=V'_{i}\left(\tau\right)$, where
\begin{equation}
\dot{\phi}_{i}\left(\tau\right)\equiv\frac{\partial\phi_{i}\left(\tau\right)}{\partial\tau}=e^{-\phi_{i}\left(\tau\right)}\frac{1}{i}\frac{\partial}{\partial\tau}e^{\phi_{i}\left(\tau\right)}.\end{equation}
The quantity $\phi\left(\tau\right)$ is the $\mathrm{U}\left(1\right)$
\textit{phase} field and satisfies the periodicity condition $\phi_{i}\left(\beta\right)=\phi_{i}\left(0\right)$
as a consequence of the periodic properties of the $V'_{i}\left(\tau\right)$
field in Eq. (\ref{periodic part V}). Next, we perform the local
gauge transformation to the new bosonic variables\begin{equation}
\left[\begin{array}{c}
b\left(\tau\right)\\
\bar{b}_{i}\left(\tau\right)\end{array}\right]=\left[\begin{array}{cc}
e^{i\phi_{i}\left(\tau\right)} & 0\\
0 & e^{-i\phi_{i}\left(\tau\right)}\end{array}\right]\left[\begin{array}{c}
a_{i}\left(\tau\right)\\
\bar{a}_{i}\left(\tau\right)\end{array}\right]\end{equation}
that removes the imaginary term $-i\int_{0}^{\beta}d\tau\dot{\phi}_{i}\left(\tau\right)n_{bi}\left(\tau\right)$
from all the Fourier modes except at zero frequency. From the above
we deduce bosons have a composite nature made of bosonic part $a_{i}\left(\tau\right)$
and attached {}``flux'' $\exp\left[i\phi_{i}\left(\tau\right)\right]$.
Due to such $\mathrm{U}\left(1\right)$ gauge invariance, the fluctuations
and the phase have the dynamics of $\mathrm{U}\left(1\right)$ gauge
field.

\subsection{Gauge group $\mathrm{U\left(1\right)}$ governed phase only action}

By integrating out the auxiliary static field $V_{i0}$ we calculate
the partition function with an effective action expressed in the form
of the propagator $\hat{G}$\begin{equation}
\mathcal{Z}=\int\left[\mathcal{D}\phi\right]e^{-\sum_{i}\int_{0}^{\beta}d\tau\left[\frac{1}{2U_{\mathrm{eff}}}\dot{\phi}_{i}^{2}\left(\tau\right)+\frac{1}{i}\frac{\bar{\mu}_{b}}{U_{\mathrm{eff}}}\dot{\phi}_{i}\left(\tau\right)\right]+\mathrm{Tr}\ln\hat{G}^{-1}},\label{partition function propagator}\end{equation}
where $\bar{\mu}_{b}/U_{b}=\mu_{b}/U_{b}+1/2$ is the shifted reduced
bosonic chemical potential. In the above $\exp\left(-\mathrm{Tr}\ln\hat{G}^{-1}\right)\equiv\det\hat{G}$
and the determinant takes the form\begin{eqnarray}
\det\hat{G} & = & \int\left[\mathcal{D}\bar{a}\mathcal{D}a\right]\exp\left\{ -\sum_{\left\langle i,j\right\rangle }\int_{0}^{\beta}d\tau\right.\nonumber \\
 & \times & \bar{a}_{i}\left(\tau\right)\left[\left(\frac{\partial}{\partial\tau}+\bar{\mu}_{b}\right)\delta_{ij}\right.\nonumber \\
 & - & \left.\left.e^{i\phi_{i}\left(\tau\right)}t_{bij}e^{-i\phi_{j}\left(\tau\right)}\right]a_{i}\left(\tau\right)\right\} .\end{eqnarray}
We parametrize the boson fields $a_{i}\left(\tau\right)=a_{0}+a_{i}^{'}\left(\tau\right)$
and incorporate fully our calculations to the phase fluctuations governed
by the gauge group $\mathrm{U}\left(1\right)$. Assuming nonfluctuating
amplitude at low temperatures $a_{i}\left(\tau\right)=a_{0}$, we
drop the corrections, which was proved to be justified in the large
$U_{b}/t_{b}$ limit we are interested in \cite{polak,kampf}. The
amplitude fluctuations are massive one and do not play important role
in the low energy scales. It is very convenient to define the order
parameter \begin{equation}
\Psi_{\mathrm{B}}\equiv\left\langle b_{i}\left(\tau\right)\right\rangle =\left\langle a_{i}\left(\tau\right)\exp\left[i\phi_{i}\left(\tau\right)\right]\right\rangle =a_{0}\psi_{\mathrm{B}},\label{order parameter definition}\end{equation}
which signals the emergence of the superfluid phase and vanishes in
the Mott-insulator state. The SF state is characterized by spontaneously
breaking of the $\mathrm{U\left(1\right)}$ symmetry of Bose-Fermi-Hubbard
Hamiltonian. Note, that a nonzero value of the amplitude $a_{0}$
in Eq. (\ref{order parameter definition}) is \emph{not sufficient}
for superfluidity. To achieve this, also the phase variables $\phi$
in Eq. (\ref{order parameter definition}), must become stiff and
coherent, which implies $\psi_{\mathrm{B}}\neq0$. As we see in the
next sections the presence of the fermions and density-density interactions
$U_{bf}$ between species of different statistics can also change
the amplitude of the order parameter. After mentioned assumption the
inverse of the propagator becomes\begin{equation}
\hat{G}^{-1}=\hat{G}_{0}^{-1}-\hat{T}=\hat{G}_{0}^{-1}\left(1-\hat{T}\hat{G}_{0}\right).\end{equation}
The explicit value of the amplitude $a_{0}$ in Eq. (\ref{order parameter definition})
can be obtained from minimization of the Hamiltonian $\partial\mathcal{H}\left(a_{0}\right)/\partial a_{0}=0$.
Therefore, we write\begin{eqnarray}
\hat{G}_{0} & = & a_{0}^{2}\equiv\frac{\sum_{\left\langle i,j\right\rangle }t_{bij}+\bar{\mu}_{b}}{U_{b}}-\frac{U_{bf}}{U_{b}}n_{\mathrm{F}}.\label{zero mode}\\
\hat{T} & = & e^{i\phi_{i}\left(\tau\right)}t_{bij}e^{-i\phi_{j}\left(\tau\right)}.\label{hopping}\end{eqnarray}
Expanding the trace of the logarithm in Eq. (\ref{partition function propagator})
and making use the above we obtain up to the second order in the amplitude
of the order parameter Eq. (\ref{order parameter definition}) \begin{eqnarray}
\mathrm{Tr}\ln\hat{G}^{-1} & = & -\mathrm{Tr}\ln\hat{G}_{0}-\mathrm{Tr}\hat{T}\hat{G}_{0}\nonumber \\
 & - & \frac{1}{2}\mathrm{Tr}\left(\hat{T}\hat{G}_{0}\right)^{2}.\end{eqnarray}
Trace over first term of the expansion, as previously, not containing
any fluctuating field variables, gives an inessential constant contribution
to the action. Let us consider the second order term in more detail\begin{eqnarray}
\mathrm{Tr}\left(\hat{T}\hat{G}_{0}\right) & = & \sum_{\left\langle i,j\right\rangle }\tilde{t}_{bij}\int_{0}^{\beta}d\tau d\tau'\nonumber \\
 & \times & e^{-i\left[\phi_{i}\left(\tau\right)-\phi_{j}\left(\tau'\right)\right]}\delta\left(\tau-\tau'\right),\end{eqnarray}
where\begin{equation}
\tilde{t}_{bij}=a_{0}^{2}t_{bij}=\left(\frac{\sum_{\left\langle i,j\right\rangle }t_{bij}+\bar{\mu}_{b}}{U_{b}}-\frac{U_{bf}}{U_{b}}n_{f}\right)t_{bij},\end{equation}
the hopping matrix elements are re-normalized by the amplitude of
the order parameter. We see that in comparison to pure bosonic case
there is an additional shift $-U_{bf}n_{\mathrm{F}}/U_{b}$ that depends
on the average of the fermion concentration and normalized inter-species
interaction. The above was also obtained in the effective bosonic
model and recognized as a mean-field contribution \cite{mering}.
Finally, the partition function Eq. (\ref{partition function propagator})
becomes\begin{eqnarray}
\mathcal{Z} & = & \int\left[\mathcal{D}\phi\right]e^{-\mathcal{S}_{\mathrm{phase}}\left[\phi\right]}\end{eqnarray}
with an effective action expressed \emph{only} in the \emph{phase}
fields variable\begin{eqnarray}
\mathcal{S}_{\mathrm{phase}}\left[\phi\right] & = & \int_{0}^{\beta}d\tau\left\{ \sum_{i}\left[\frac{1}{2U_{\mathrm{eff}}}\dot{\phi_{i}^{2}}\left(\tau\right)+\frac{1}{i}\frac{\bar{\mu}_{b}}{U_{\mathrm{eff}}}\dot{\phi_{i}}\left(\tau\right)\right]\right.\nonumber \\
 & - & \left.\tilde{t}_{b}\sum_{i,j}e^{\phi_{i}\left(\tau\right)}I_{ij}e^{-\phi_{j}\left(\tau\right)}\right\} ,\label{action only phase}\end{eqnarray}
where $\tilde{t}_{b}=\left(\sum_{\left\langle i,j\right\rangle }t_{bij}+\bar{\mu}_{b}-U_{bf}n_{\mathrm{F}}\right)t_{b}/U_{b}$.
The total time derivative Berry phase imaginary term in Eq. (\ref{action only phase})
is nonzero due to topological phase field configurations with $\phi_{i}\left(\beta\right)-\phi_{i}\left(0\right)=2\pi m_{i}$
($m_{i}=0,\pm1,\pm2...$) that results in topological ingredients
to the correlator we will see below. Therefore, we concentrate on
closed paths in the {}``imaginary time'' $\left(0,\beta\right)$
labeled by the integer winding numbers $m_{i}$. The path-integral
\begin{equation}
\int\left[\mathcal{D}\phi\right]...\equiv\sum_{\left[m_{i}\right]}\int_{0}^{2\pi}\left[\mathcal{D}\phi\left(0\right)\right]\int_{_{\phi_{i}\left(0\right)}}^{\phi_{i}\left(\tau\right)+2\pi m_{i}}\left[\mathcal{D}\phi\left(\tau\right)\right]...,\end{equation}
includes a summation over $m_{i}$ and in each topological sector
the integration goes over the gauge potentials. Therefore, we do not
ignore the compactness of the gauge fields.

To proceed, we replace the phase degrees of freedom by the uni-modular
scalar complex field $\psi$ which satisfies the quantum periodic
boundary condition $\psi_{i}\left(\beta\right)=\psi_{i}\left(0\right)$.
This can be conveniently done using the Fadeev-Popov method with Dirac
delta functional resolution of unity, where we take $\psi$ as a continuous
but constrained (on the average) variable to have the uni-modular
value. We introduce \begin{eqnarray}
1 & = & \int\left[\mathcal{D}\psi\mathcal{D}\psi^{*}\right]\delta\left(\sum_{i}\left|\psi\left(\tau\right)\right|^{2}-N\right)\nonumber \\
 & \times & \delta\left(\psi_{i}-e^{i\phi_{i}\left(\tau\right)}\right)\delta\left(\psi_{i}^{*}-e^{-i\phi_{i}\left(\tau\right)}\right)\end{eqnarray}
and\begin{eqnarray}
\delta\left(\sum_{i}\left|\psi_{i}\left(\tau\right)\right|^{2}-N\right) & = & \frac{1}{2\pi i}\int_{-i\infty}^{+i\infty}d\lambda\nonumber \\
 & \times & e^{\int_{0}^{\beta}d\tau\lambda\left(\sum_{i}\left|\psi_{i}\left(\tau\right)\right|^{2}-N\right)},\end{eqnarray}
where $N$ is the number of lattice sites. Introducing the Lagrange
multiplier $\lambda$, which adds the quadratic terms (in the $\psi$
fields) to the action we can solve for the constraint. The partition
function can be rewritten to the form\begin{eqnarray}
\mathcal{Z} & = & \frac{1}{2\pi i}\int_{-i\infty}^{+i\infty}e^{-\lambda N}d\lambda\int\left[\mathcal{D}\psi\mathcal{D}\psi^{*}\right]\nonumber \\
 & \times & \exp\left\{ -\sum_{i,j}\int_{0}^{\beta}d\tau d\tau'\psi_{i}\left[\left(\tilde{t}_{b}I_{ij}+\lambda\delta_{ij}\right)\delta\left(\tau-\tau'\right)\right.\right.\nonumber \\
 & + & \left.\left.\gamma_{ij}\left(\tau,\tau'\right)\right]\psi_{j}^{*}\right\} ,\end{eqnarray}
where \begin{equation}
\gamma_{ij}\left(\tau,\tau'\right)=\left\langle \exp\left\{ -i\left[\phi_{i}\left(\tau\right)-\phi_{j}\left(\tau'\right)\right]\right\} \right\rangle \end{equation}
is the two-point phase correlator associated with the order parameter
field, where $\left\langle \cdots\right\rangle $ denotes averaging
with respect to the action in Eq. (\ref{action only phase}). Because
the values of the phases $\phi$ which differ by $2\pi$ are equivalent
we decompose phase field in terms of a periodic field and term linear
in $\tau$:\begin{equation}
\phi_{i}\left(\tau\right)=\varphi_{i}\left(\tau\right)+\frac{2\pi}{\beta}m_{i}\tau\end{equation}
with $\phi_{i}\left(\beta\right)=\phi_{i}\left(0\right).$ As a result
the phase correlator factorizes as the product of a topological term
depending on the integers $m_{i}$ and non-topological one:\begin{equation}
\gamma_{ij}\left(\tau,\tau'\right)=\gamma_{ij}^{T}\left(\tau,\tau'\right)\gamma_{ij}^{N}\left(\tau,\tau'\right).\end{equation}
Performing the Poisson re-summation formula in \begin{equation}
\gamma_{ij}^{T}\left(\tau,\tau'\right)=\frac{\sum_{\left[m_{i}\right]}e^{-i\frac{2\pi}{\beta}\left(\tau-\tau'\right)m_{i}}e^{-\frac{2\pi}{\beta}\sum_{i}\left[\frac{\pi}{U_{\mathrm{eff}}}m_{i}^{2}+\frac{\beta}{i}\frac{\bar{\mu}_{b}}{U_{\mathrm{eff}}}m_{i}\right]}}{\sum_{\left[m_{i}\right]}e^{-\frac{2\pi}{\beta}\sum_{i}\left[\frac{\pi}{U_{\mathrm{eff}}}m_{i}^{2}+\frac{\beta}{i}\frac{\bar{\mu}_{b}}{U_{\mathrm{eff}}}m_{i}\right]}}\end{equation}
and the functional integration over the phase variables \begin{equation}
\gamma_{ij}^{N}\left(\tau,\tau'\right)=\frac{\int\left[\mathcal{D}\varphi\right]e^{-i\left[\varphi_{i}\left(\tau\right)-\varphi_{j}\left(\tau'\right)\right]}e^{-\sum_{i}\frac{1}{2U_{\mathrm{eff}}}\int_{0}^{\beta}d\tau\dot{\varphi_{i}^{2}}\left(\tau\right)}}{\int\left[\mathcal{D}\varphi\right]e^{-\sum_{i}\frac{1}{2U_{\mathrm{eff}}}\int_{0}^{\beta}d\tau\dot{\varphi_{i}^{2}}\left(\tau\right)}}\end{equation}
the final formula of the correlator takes the form \begin{eqnarray}
\gamma_{ij}\left(\tau,\tau'\right) & = & \frac{\vartheta\left(\pi\frac{\bar{\mu}_{b}}{U_{\mathrm{eff}}}+\pi\frac{\tau-\tau'}{\beta},e^{-\frac{1}{U_{\mathrm{eff}}}\frac{2\pi^{2}}{\beta}}\right)}{\vartheta\left(\pi\frac{\bar{\mu}_{b}}{U_{\mathrm{eff}}},e^{-\frac{1}{U_{\mathrm{eff}}}\frac{2\pi^{2}}{\beta}}\right)}\nonumber \\
 & \times & \exp\left(\frac{U_{\mathrm{eff}}}{2}\left|\tau-\tau^{'}\right|-\frac{\left(\tau-\tau'\right)^{2}}{\beta}\right),\end{eqnarray}
where $\vartheta\left(z,q\right)$ is the Jacobi theta function, which
comes from the topological contribution - summation over integer winding
numbers. The function $\vartheta\left(z,q\right)$ is defined by \begin{equation}
\vartheta\left(z,q\right)=1+2\sum_{n=1}^{+\infty}\cos\left(2nz\right)q^{n^{2}}\end{equation}
and is $\beta$-periodic in the {}``imaginary time'' as well in
the variable $\bar{\mu}_{b}/U_{\mathrm{eff}}$ with the period of
unity which emphasizes the special role of its integer values. After
Fourier transforming one obtains

\begin{equation}
\gamma_{ij}\left(\omega_{\nu}\right)=\frac{1}{\mathcal{Z}_{0}}\frac{4}{U_{\mathrm{eff}}}\sum_{\left[m_{i}\right]}\frac{e^{-\frac{U_{\mathrm{eff}}\beta}{2}\sum_{i}\left(m_{i}+\frac{\bar{\mu}_{b}}{U_{\mathrm{eff}}}\right)^{2}}}{1-4\left(\sum_{i}m_{i}+\frac{\bar{\mu}_{b}}{U_{\mathrm{eff}}}-i\frac{\omega_{\ell}}{U_{\mathrm{eff}}}\right)^{2}},\label{topological contribution}\end{equation}
where\begin{equation}
\mathcal{Z}_{0}=\sum_{\left[m_{i}\right]}e^{-\frac{U_{\mathrm{eff}}\beta}{2}\sum_{i}\left(m_{i}+\frac{\bar{\mu}_{b}}{U_{\mathrm{eff}}}\right)^{2}}\end{equation}
is the partition function for the set of quantum rotors. The action
Eq. (\ref{action only phase}), with the topological contribution
Eq. (\ref{topological contribution}), after Fourier transform, is
written as \begin{equation}
\mathcal{S}_{\mathrm{eff}}\left[\psi,\bar{\psi}\right]=\frac{1}{N\beta}\sum_{\mathbf{k},\ell}\bar{\psi}_{\mathbf{k}}\left(\omega_{\ell}\right)\mathrm{\Gamma}_{\mathbf{k}}^{-1}\left(\omega_{\ell}\right)\psi_{\mathbf{k}}\left(\omega_{\ell}\right),\end{equation}
where $\mathrm{\Gamma}_{\mathbf{k}}^{-1}\left(\omega_{\ell}\right)=\lambda-t_{b\mathbf{k}}+\gamma^{-1}\left(\omega_{\ell}\right)$
is the inverse of the propagator and $t_{b\boldsymbol{k}}$ is the
Fourier transform of the bosonic hopping matrix elements for two-
$t_{b\boldsymbol{k}}^{2D}=2t_{b}\left(\cos k_{x}+\cos k_{y}\right)$
and three-dimensional $t_{b\boldsymbol{k}}^{3D}=2t_{b}\left(\cos k_{x}+\cos k_{y}+\cos k_{z}\right)$
lattice.

\section{Critical Line}

Within the phase coherent state the order parameter $\psi_{B}$ is
evaluated in the thermodynamic limit $N\rightarrow\infty$ by the
saddle point method $\delta\mathcal{F}/\delta\lambda=0$ and the uni-modular
condition of the $\mathrm{U}\left(1\right)$ phase variables translates
into the equation\begin{equation}
1-\psi_{\mathrm{B}}^{2}=\lim_{N\rightarrow\infty}\frac{1}{N\beta}\sum_{\mathbf{k},\ell}\Gamma_{\boldsymbol{k}}\left(\omega_{\ell}\right),\label{critical line}\end{equation}
with\begin{equation}
\Gamma_{\boldsymbol{k}}^{-1}\left(\omega_{\ell}\right)=\tilde{t}_{b\mathbf{k}=0}-\tilde{t}_{b\mathbf{k}}+\frac{1}{U_{\mathrm{eff}}}\bar{\mu}_{b}^{2}-\frac{1}{U_{\mathrm{eff}}}\left(\bar{\mu}_{b}-i\omega_{\ell}\right)^{2}.\end{equation}
The phase boundary is determined by the divergence of the order parameter
susceptibility $\Gamma_{\mathbf{k}=0}\left(\omega_{\ell=0}\right)=0$\begin{equation}
\lambda_{0}-t_{b\boldsymbol{k}=0}^{\mathrm{max}}+\gamma^{-1}\left(\omega_{\ell=0}\right)=0\label{lagrange 0}\end{equation}
which determines the critical value of the Lagrange parameter $\lambda=\lambda_{0}$
and stays constant in the whole global coherent phase. To proceed,
it is desirable to introduce the density of states \begin{equation}
\rho\left(\xi\right)=\frac{1}{N}\sum_{\boldsymbol{k}}\delta\left(\xi-\frac{t_{b\boldsymbol{k}}}{t_{b}}\right)\label{DOS definition}\end{equation}
because the analytical expressions we use can be advantageous in evaluating
sums over momenta. The corresponding formulas for square lattice can
be written as\begin{equation}
\rho^{2D}\left(\xi\right)=\frac{1}{2\pi^{2}t_{b}}\mathbf{K}\left(\sqrt{1-\left(\frac{\xi}{4t_{b}}\right)^{2}}\right)\Theta\left(1-\left|\frac{\xi}{4t_{b}}\right|\right),\end{equation}
and for simple cubic geometry takes form \begin{eqnarray}
\rho^{3D}\left(\xi\right) & = & \frac{1}{2\pi^{3}t_{b}}\int_{a_{1}}^{a_{2}}\frac{d\epsilon}{\sqrt{1-\epsilon^{2}}}\Theta\left(1-\frac{\left|\xi\right|}{6t_{b}}\right)\nonumber \\
 & \times & \mathbf{K}\left(\sqrt{1-\left(\frac{\xi}{4t_{b}}+\epsilon\right)^{2}}\right)\end{eqnarray}
with $a_{1}=\mathrm{min}\left(-1,-2-\xi/2t_{b}\right)$ and $a_{2}=\mathrm{max}\left(1,2-\xi/2t_{b}\right)$;
$\mathbf{K}\left(x\right)$ is the elliptic function of the first
kind.\cite{abramovitz}. After summation over Bose-Matsubara frequency
and for zero temperature limit $\beta\rightarrow\infty$ we can rewrite
the critical line equation to the form that represents solution of
the BF Hubbard model in terms of re-normalized pure Bose-Hubbard Hamiltonian
in the quantum rotor approach:

\begin{widetext}\begin{eqnarray}
1-\psi_{\mathrm{B}}^{2} & = & \frac{1}{2}\int_{-\infty}^{+\infty}\frac{\rho\left(\xi\right)d\xi}{\sqrt{2\left(\xi_{\mathrm{max}}-\xi\right)\left(2z\frac{t_{b}}{U_{b}}+\frac{\mu_{b}}{U_{b}}-\eta+\frac{1}{2}\right)\frac{1}{\alpha}\frac{t_{b}}{U_{b}}+\upsilon^{2}\left(\frac{1}{\alpha}\frac{\mu_{b}}{U_{b}}\right)}}\label{critical line-1}\end{eqnarray}

\end{widetext}In the above $\upsilon\left(\mu_{b}/\alpha U_{b}\right)=\mathrm{frac}\left(\mu_{b}/\alpha U_{b}\right)-1/2,$
where $\mathrm{frac}\left(x\right)=x-\left[x\right]$ is the fractional
part of the number and $\left[x\right]$ is the floor function which
gives the greatest integer less than or equal to $x$; $\xi_{\mathrm{max}}$
stands for the maximum value of the bosonic dispersion spectrum $t_{b\boldsymbol{k}}$
and $z$ is the lattice coordination number. The renormalization parameters
are defined as: \begin{eqnarray}
\alpha & = & 1+\frac{U_{bf}^{2}}{U_{b}}\chi'\left(\omega,\mu_{f}\right)\label{parameters1}\\
\eta & = & \frac{U_{bf}}{U_{b}}n_{\mathrm{F}}\label{parameters2}\end{eqnarray}
and allow us to see how adding free fermions to strongly interacting
bosons confined in optical lattice influences the phase boundary.

\section{Phase Diagrams - BH model}

The zero-temperature phase diagram of the Bose-Fermi-Hubbard model
Eq. (\ref{hamiltonian}) can be calculated from Eq. (\ref{critical line-1})
and usually is plotted as a function of $t_{b}/U_{b}$, with the density
of the bosons controlled by a chemical potential $\mu_{b}/U_{b}$.
The presence of the fermions implicates two additional different parameters
that can by varied namely $\alpha$ and $\eta$ in Eq. (\ref{parameters1})
and Eq. (\ref{parameters2}). The strength of the inter-species interaction
influences both of them, however the sign of $U_{bf}$ and the average
density of fermions $n_{\mathrm{F}}$ added to the system affects
only $\eta$. Besides, in the local dynamic approach the sign of the
density-density interaction depends also on the normalized frequency
$\omega/t_{f}$. For a general choice of parameters, Eq. (\ref{critical line-1})
is easy to solve, however considerations of special cases can provide
more insights into the solution of the problem. In discussion we will
follow the scheme\begin{eqnarray}
\eta\begin{array}{c}
\nearrow\\
\searrow\end{array} & \begin{array}{cccc}
\alpha<1 & \mbox{and} & \begin{array}{c}
c_{b}/v_{f}>1\\
c_{b}/v_{f}\thicksim1\end{array} & \begin{array}{c}
\mbox{for \ensuremath{\omega=0}}\\
\mbox{for \ensuremath{\omega<\omega_{\mathrm{crit}}}}\end{array}\\
\\\alpha>1 & \mbox{and} & c_{b}/v_{f}\thicksim1 & \mbox{for \ensuremath{\omega>\omega_{\mathrm{crit}}}}\end{array}\end{eqnarray}
firstly choosing the sign of the $\eta$ and later $\alpha$ in the
static ($\omega=0$) or dynamic ($\omega\neq0$) limit (see Fig. \ref{fig1}). 

Before we proceed with analysis let us introduce the notation for
the maximum of the critical value for parameter $t_{b}/U_{b}$ (as
a function of the normalized chemical potential $\mu_{b}/U_{b}$)
at the tip of the $n$th ($n_{\mathrm{B}}=n$) MI lobe for different
lattice geometries and model parameters $\alpha$ and $\eta$ as follows
\begin{equation}
x_{n}\left(\alpha,\eta\right)\equiv{\rm max}\left\{ \left(\frac{t_{b}}{U_{b}}\right)_{{\rm crit}}\right\} _{\alpha,\eta}^{2D,3D}.\end{equation}
\begin{figure}
\includegraphics[scale=0.85]{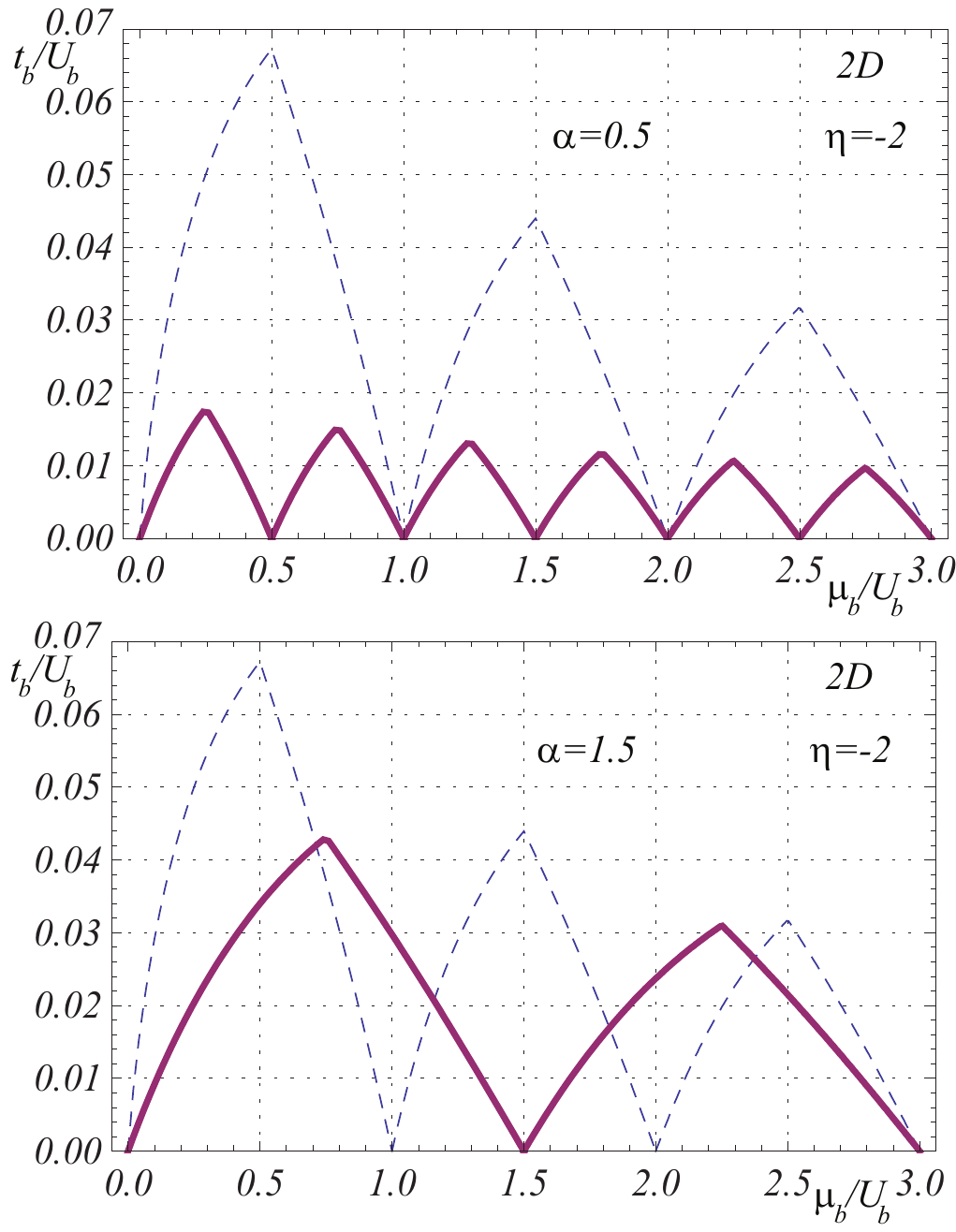}\caption{(Color online) Phase diagrams ($t_{b}/U_{b}$-$\mu_{b}/U_{b}$) for
the square ($2D$) lattice for different $\alpha=0.5$ (higher panel,
$\omega<\omega_{\mathrm{crit}}$), $\alpha=1.5$ ($\omega>\omega_{\mathrm{crit}}$)
and $\eta=-2$ (negative scattering length $a_{bf}<0$). Dashed line
stands for the phase boundary of one-component Bose-Hubbard model.
Within the lobes the Mott insulator phase takes place with $\Psi_{\mathrm{B}}=0$.\label{fig4}}

\end{figure}
\begin{figure}
\includegraphics[scale=0.85]{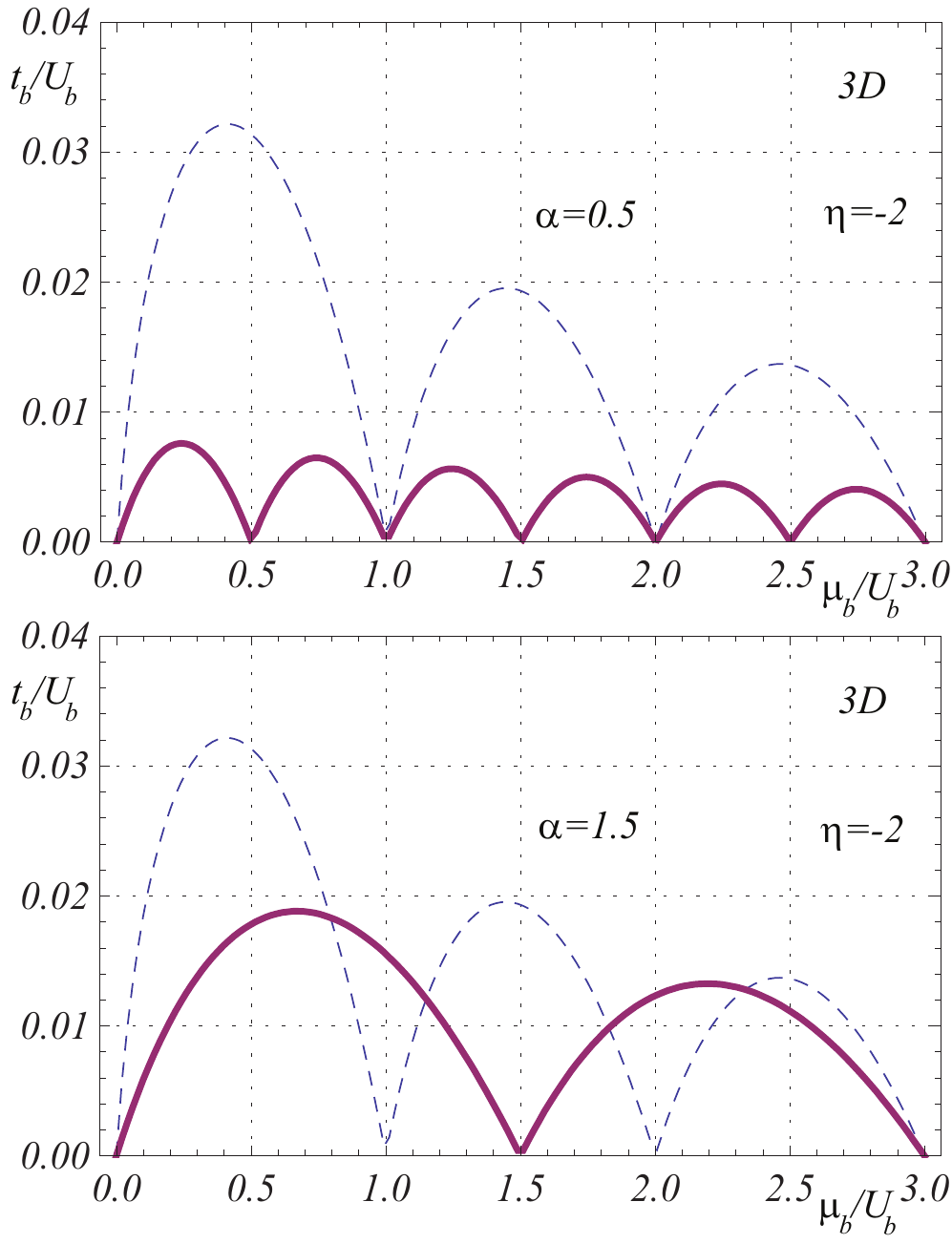}\caption{(Color online) Phase diagrams ($t_{b}/U_{b}$-$\mu_{b}/U_{b}$) for
the cubic ($3D$) lattice for different $\alpha=0.5$ (higher panel,
$\omega<\omega_{\mathrm{crit}}$), $\alpha=1.5$ ($\omega>\omega_{\mathrm{crit}}$)
and $\eta=-2$ (negative scattering length $a_{bf}<0$). Dashed line
stands for the phase boundary of one-component Bose-Hubbard model.
Within the lobes the Mott insulator phase takes place with $\Psi_{\mathrm{B}}=0$.\label{fig5}}

\end{figure}
The above determines when the transition from MI to SF occurs. Values
$\alpha=1$ and $\eta=0$ stand for the one-component bosonic case.
In Table \ref{comparison} we show comparison of $x_{n}\left(1,0\right)$
for higher densities of the particles calculated in the quantum rotor
approach (QRA) to very accurate, recently developed, diagrammatic
perturbation approach\cite{teichmann} to Bose-Hubbard Hamiltonian.
The results for $3D$ BH model obtained in both theories are very
close and also comparison to quantum Monte-Carlo (QMC) numerical calculations\cite{capogrosso-sansone}
indicates that methods we use are able to properly catch the interesting
physics of strongly interacting systems. However, we want to analyze
the phase boundary for number of particles per lattice sites higher
than one $n_{\mathrm{B}}>1$ that adds another dimension to the analysis
and is difficult for the QMC to catch. The phase boundary for square
lattice shows that QRA works well also in low-dimensional geometries,
especially for higher densities. Nevertheless, the structure of Eq.
(\ref{critical line-1}) can cause some problems when \emph{both}
$\alpha$ and $\eta$ are nonzero and the number of bosons is equal
one per lattice sites. We expect that for $\eta>1$ some artificial
effects may arise for values of the normalized chemical potential
$\mu_{b}/U_{b}\approx0$ close to zero.

\begin{table}
\begin{tabular}{c|c|c|c|c|c}
$2D$ & $n_{\mathrm{B}}=1$ & $2$ & $3$ & $4$ & $10$\tabularnewline
\hline 
DPT & $0.0590934$ & $0.0348009$ & $0.0247350$ & $0.0191986$ & $0.0082079$\tabularnewline
\hline 
QRA & $0.0671998$ & $0.0439387$ & $0.0317523$ & $0.0246185$ & $0.0093296$\tabularnewline
\hline 
$3D$ & $n_{\mathrm{B}}=1$ & $2$ & $3$ & $4$ & $10$\tabularnewline
\hline 
DPT & $0.0340685$ & $0.0200755$ & $0.0142709$ & $0.0110779$ & $0.0047362$\tabularnewline
\hline 
QRA & $0.0321429$ & $0.0194846$ & $0.0136102$ & $0.0103755$ & $0.0042086$\tabularnewline
\end{tabular}\caption{Comparison of the maximum of the critical value for $\left(t_{b}/U_{b}\right)_{\mathrm{crit}}$
(as a function of the normalized bosonic chemical potential $\mu_{b}/U_{b}$)
at the tip of the $n$th ($n_{\mathrm{B}}=1\div4$ and $10$) Mott
insulator lobe for the square ($2D$) and cubic ($3D$) lattice in
the one-component Bose-Hubbard model: DPT - diagrammatic perturbation
theory\cite{teichmann}), QRA - our calculations using quantum rotor
approach).}
\label{comparison}
\end{table}

\section{Bose-Fermi-Hubbard phase diagram}

\begin{figure}
\includegraphics[scale=0.8]{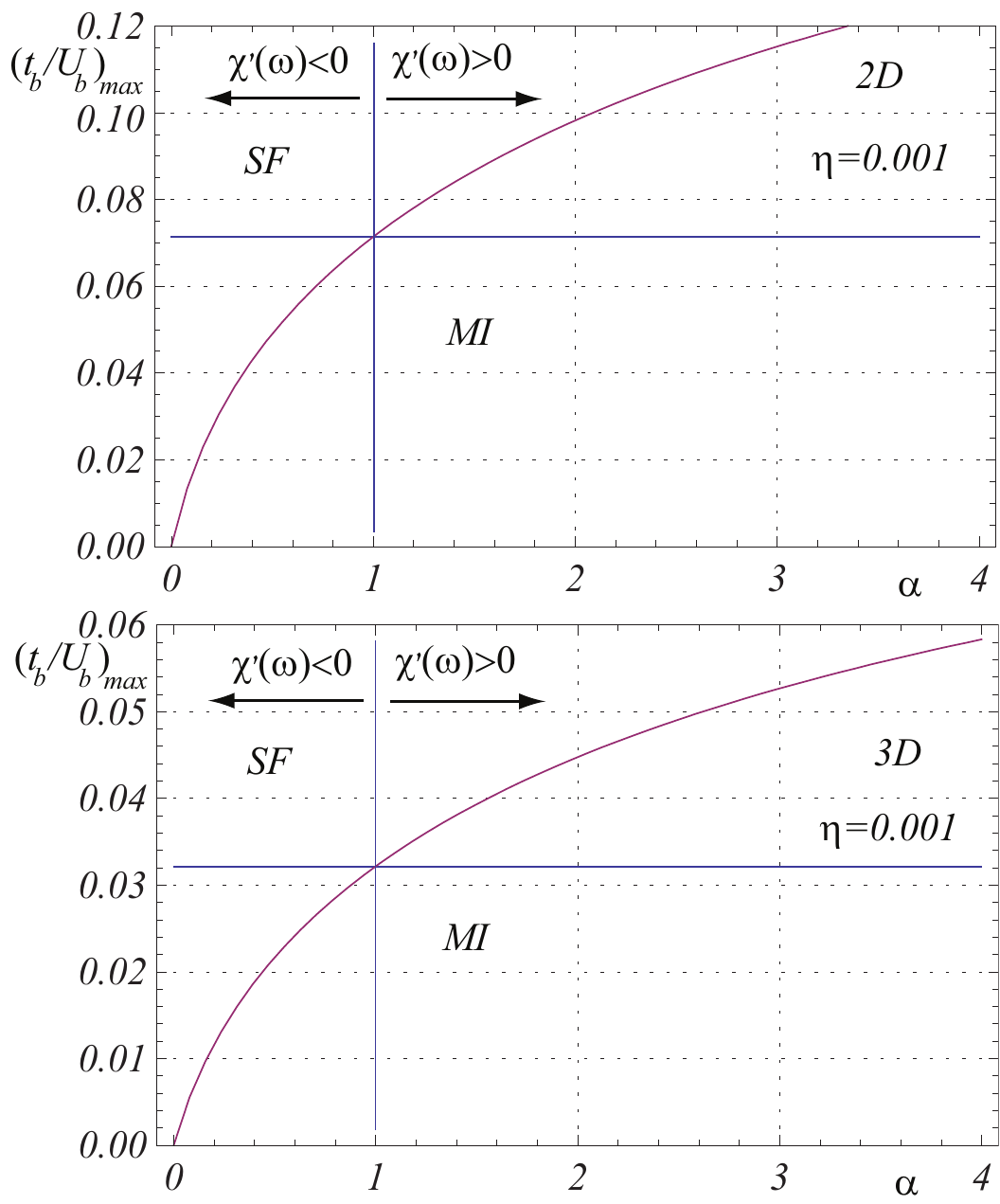}\caption{(Color online) Maximum value of the $\left(t_{b}/U_{b}\right)_{\mathrm{max}}$
for different $\alpha$ and $\eta=0.001$ (very small amount of fermions)
for square ($2D$) and cubic ($3D$) lattice with $n_{\mathrm{B}}=1$.
Vertical solid line stands for the $\left(t_{b}/U_{b}\right)_{\mathrm{max}}$
obtained in one-component Bose-Hubbard model with one particle per
lattice site. Above the curves the superfluid phase takes place with
$\Psi_{\mathrm{B}}\neq0$.\label{fig6}}

\end{figure}

In the experiments\cite{best} for a degenerate mixtures of $4\times10^{5}$
$^{87}\mathrm{Rb}$ bosons and $3\times10^{5}$ $^{40}\mathrm{K}$
fermions the scattering length $a_{bf}$ (and in consequence interaction
$U_{bf}$, see Eq. (\ref{ubfexp})) can be continuously tune between
$-170a_{0}\div+800a_{0}$ below and between $-800a_{0}\div-200a_{0}$
above Feshbach resonance, where $a_{0}$ is the Bohr radius. The form
of the parameters we choose Eq. (\ref{parameters1}) and Eq. (\ref{parameters2})
allows for its interpretation. The periodicity of the phase diagram
can be easily deduced from the periodic properties of the propagator
Eq. (\ref{topological contribution}) and strongly depends on $\alpha$
however, the interaction between species does not generate additional
Mott lobes in the phase diagram. The above is in contrast to the strong
coupling expansion and exact diagonalization method applied to the
system of two alkali-metal atoms with different masses where (for
very small lattice sizes and quenched disorder) the MI phases with
integer filing factors disappear for boson-impurity interaction energy
larger than on-site atom-atom interaction energy itself and also the
MI phase exists for incommensurate bosonic filling \cite{krutitsky}.
If we fix the number of fermions $n_{\mathrm{F}}$ and inter-species
interaction $U_{bf}$ in Eq. (\ref{parameters2}) still there is a
dynamic part of the local Lindhard function $\chi'\left(\omega,\mu_{f}\right)$
we have to take into account. In the static limit $\omega/t_{f}\rightarrow0$
(where the Lindhard response function is purely real) there is nothing
unexpected in the behavior of the critical line (see discussion below).
However, we must stress that even we left the frequency dependence
apart, there is still very interesting part of physics remained, because
the Lindhard response function for the system with regular density
of states shows logarithmic divergence as temperature goes to zero.
These singularities give rise to instabilities in the system towards
two new ground states a phase separated state or a supersolid phase
\cite{buchler,buechler1}. On the other hand the oscillation of the
induced effective interaction between bosons is the origin of the
formation of charge density waves \cite{mering}.

\begin{figure}
\includegraphics[scale=0.85]{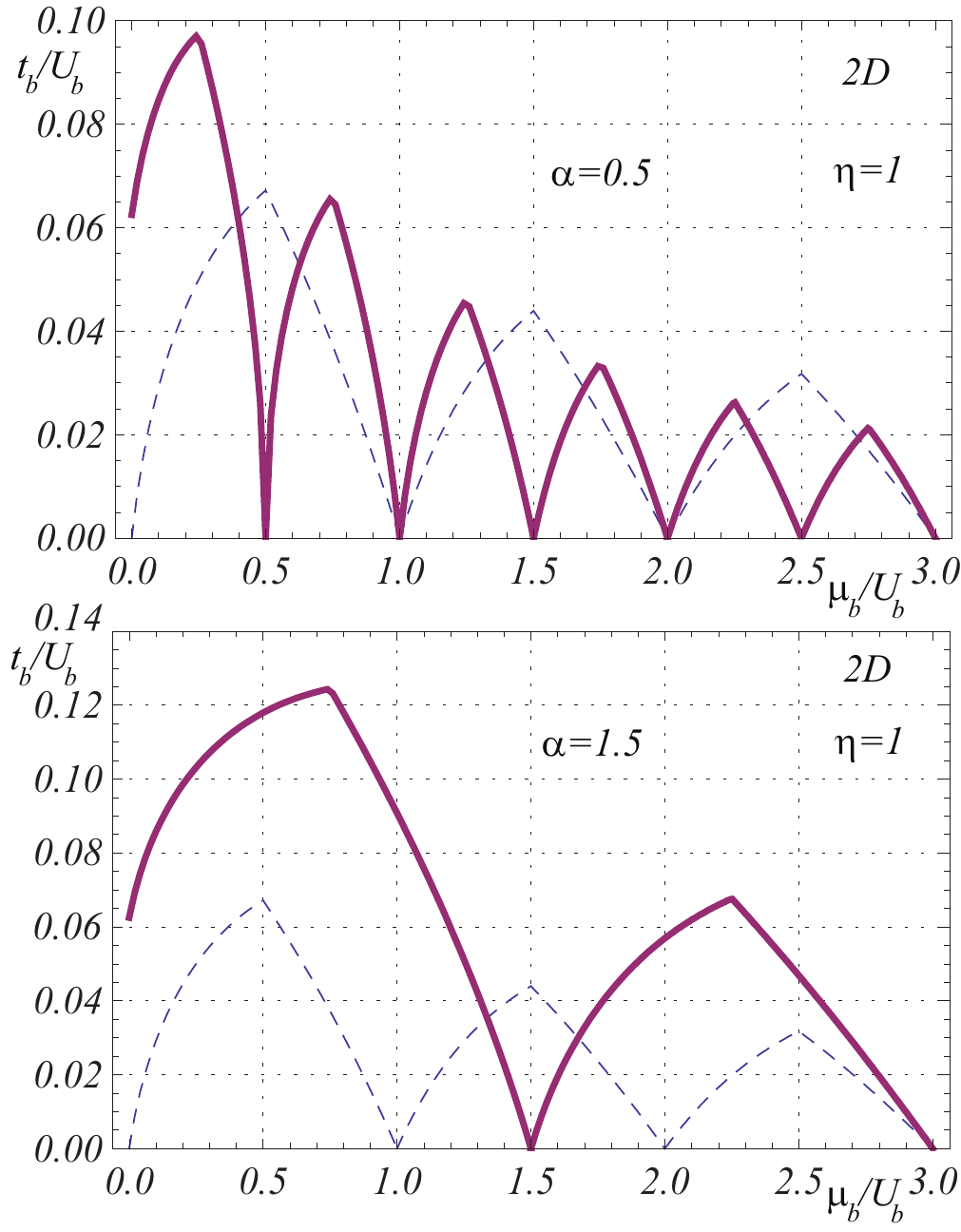}\caption{(Color online) Phase diagrams ($t_{b}/U_{b}$-$\mu_{b}/U_{b}$) for
the square ($2D$) lattice for different $\alpha=0.5$ (higher panel,
$\omega<\omega_{\mathrm{crit}}$), $\alpha=1.5$ ($\omega>\omega_{\mathrm{crit}}$)
and $\eta=1$ (positive scattering length $a_{bf}>0$). Dashed line
stands for the phase boundary of one-component Bose-Hubbard model.
Within the lobes the Mott insulator phase takes place with $\Psi_{\mathrm{B}}=0$.\label{fig7}}

\end{figure}

\begin{figure}
\includegraphics[scale=0.85]{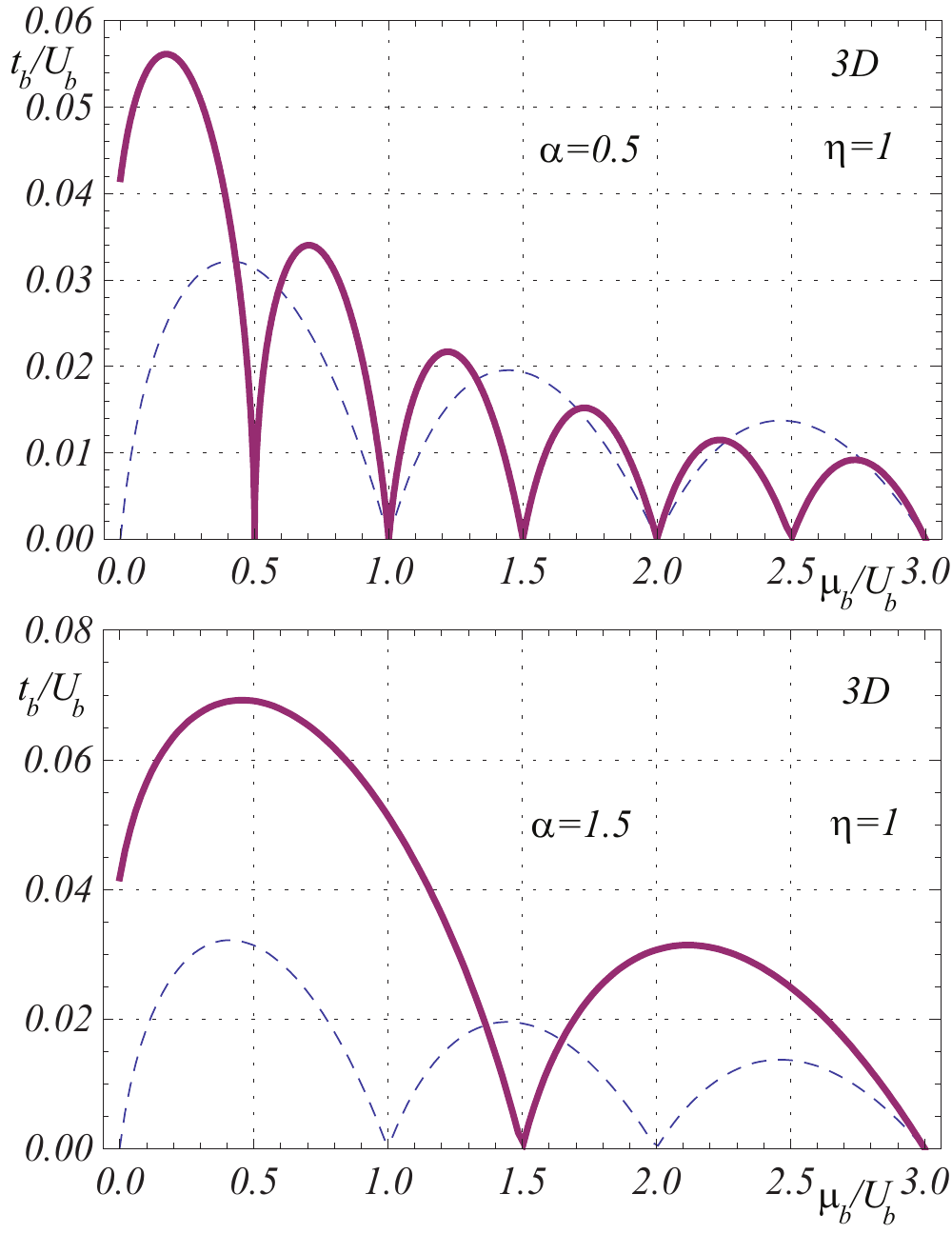}\caption{(Color online) Phase diagrams ($t_{b}/U_{b}$-$\mu_{b}/U_{b}$) for
the cubic ($3D$) lattice for different $\alpha=0.5$ (higher panel,
$\omega<\omega_{\mathrm{crit}}$), $\alpha=1.5$ ($\omega>\omega_{\mathrm{crit}}$)
and $\eta=1$ (positive scattering length $a_{bf}>0$). Dashed line
stands for the phase boundary of one-component Bose-Hubbard model.
Within the lobes the Mott insulator phase takes place with $\Psi_{\mathrm{B}}=0$.\label{fig8}}

\end{figure}

Taking $\omega<\omega_{\mathrm{crit}}$ we recover the previous theoretical
results where, after adding fermions to the system, the effective
interaction $U_{\mathrm{eff}}$ becomes smaller than repulsive energy
$U_{b}$ for bosons only (see Fig. \ref{fig4} and Fig. \ref{fig5})
and superfluid phase increases. The above is best shown on Fig. \ref{fig6}
where for very small amount of fermions $n_{\mathrm{F}}$ the parameter
$\alpha<1$ causes decreasing the Mott insulator region of the phase
diagram in comparison to the pure bosonic case. However, in the local
dynamic limit, when $\omega>\omega_{\mathrm{crit}}$ the Mott insulator
phase is becoming stronger and bosons tend to localize on the lattice
sites in both $2D$ and $3D$ cases (see Fig. \ref{fig6}, Fig. \ref{fig7}
and Fig. \ref{fig8}). %
\begin{figure}
\includegraphics[scale=0.8]{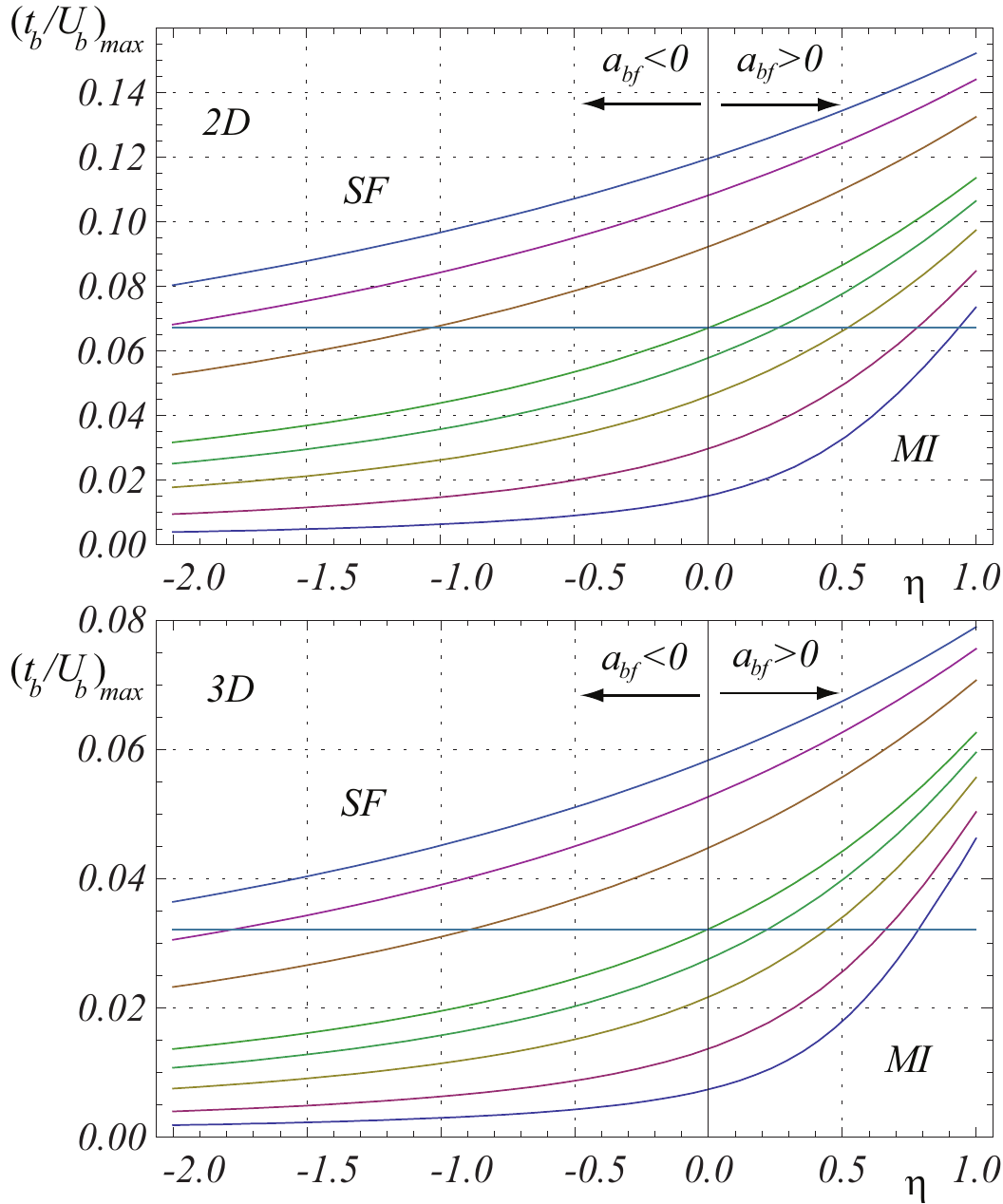}\caption{(Color online) The maximum of the critical value for the parameter
$\left(t_{b}/U_{b}\right)_{\mathrm{max}}$ with $n_{\mathrm{B}}=1$
for different $\alpha=4$, $3$, $2$, $1$, $0.75$, $0.5$, $0.25$,
$0.1$ for square ($2D$) and cubic ($3D$) lattice. Vertical solid
line stands for the $\left(t_{b}/U_{b}\right)_{\mathrm{max}}$ obtained
in the one-component Bose-Hubbard model with one particle per lattice
site. Above the curves the superfluid phase takes place with $\Psi_{\mathrm{B}}\neq0$.\label{fig9}}

\end{figure}

One may argue that the Lindhard response function depends not only
on the frequency $\omega/t_{f}$ but also on the chemical potential
$\mu_{f}/t_{f}$ for fermions and so far we did not restrict ourselves
to any particular value of it. In many approaches it is a little tricky
to handle because despite of the absence of any direct interaction
between fermions $U_{ff}=0$ the density-density fluctuations can
indeed induce some effective interaction between fermionic species
\cite{illuminati}. Therefore problem becomes complex and many theories
just take half-filled band with $n_{\mathrm{F}}=1$ so that $\mu_{f}/t_{f}=0$.
Alternative approach comes from partial particle-hole symmetry Hamiltonian
Eq. (\ref{hamiltonian}) possesses. To make our approach self-consistent
we can calculate how does a particular value of the fermionic chemical
potential change in the effectively interacting system. We remind
that the amplitude of the order parameter was obtained from minimization
condition, assuming nonfluctuating bosonic amplitude at low temperatures.
By operating a similar procedure we get a shift of the chemical potential
for fermions \begin{equation}
\mu_{f}\rightarrow\mu_{f}-U_{bf}\left(2z\frac{t_{b}}{U_{b}}+\frac{\mu_{b}}{U_{b}}-\eta+\frac{1}{2}\right),\label{shift}\end{equation}
that in the non-interacting case $U_{bf}=0$ reduces to that of free
particles obeying fermionic commutation relations (see also Appendix).

There is the limit where the system containing gaseous mixtures has
the same value of $x\left(\alpha,\eta\right)$ as in the case of only
bosons confined in optical lattice (see Fig. \ref{fig9}). Again,
we take advantage of the choice of the parameters, that suits well
our goal, and make notation of the condition very simple. If $\eta=1-\alpha$
we have\begin{equation}
x_{n}\left(1,0\right)=x_{n}\left(\alpha,1-\alpha\right).\label{condition}\end{equation}
The above, in terms of the original variables, leads to $U_{bf}n_{\mathrm{F}}/U_{b}=U_{bf}^{2}\chi'\left(\omega,\mu_{f}\right)/U_{b}$.
For cubic lattice the formula Eq. (\ref{condition}) seems to not
hold (Fig. \ref{fig9}) however is accurate with a numerical error
less than $0.17$ percent. Therefore, if the number of fermions added
to the system is equal to the inter-species interaction then bosons
behaves as if were unaffected by the presence of fermions. As a matter
of fact we have to remember about sign of the scattering length $a_{bf}$
and normalized frequency $\omega/t_{f}$ that also modifies the introduced
condition. The recent experiment\cite{best} shows that there is an
asymmetry in profiles of visibility of the interference pattern (recorded
by absorption imaging) versus the inter-species scattering length
that increases with lattice depth. Presented data indicate that visibility
shows a maximum at the position consistent with $a_{bf}=0$. Besides,
there is an asymmetry in a shift of the MI to SF transition boundary.
Our calculations can reproduce latter however if the sign of the inter-species
interaction is negative the MI phase diminishes and quite oppositely
for positive scattering length and, as we expected, some anomaly appears
at the point with $\mu_{b}/U_{b}=0$ (see Fig. \ref{fig7} and Fig.
\ref{fig8}). There is no physical reason for the phase boundary to
change a position where the chemical potential for bosons is zero
and the repulsive interactions are very strong $U_{b}\rightarrow\infty$.
Moreover, in that case the value obtained from Eq. (\ref{critical line-1})
at mentioned point ($\mu_{b}/U_{b}=0$ and fixed $\eta>0$) is constant
in whole nonzero range of the parameter $\alpha$ and depends only
on the considered topology of the system. The similar to fermion-boson
loss of coherence for the boson-boson species was found using the
Gutzwiller mean-field approach \cite{buosante}. The main effect of
the addition different species of the same statistics is that the
new structure of wedding cake appear but the oscillatory behavior
of the relevant condensate fraction does not necessarily result in
increase of the overall coherence of other species. The later is limited
exclusively to the shallow lattice depth and was never observed in
the experiments.

\begin{figure}
\includegraphics[scale=0.8]{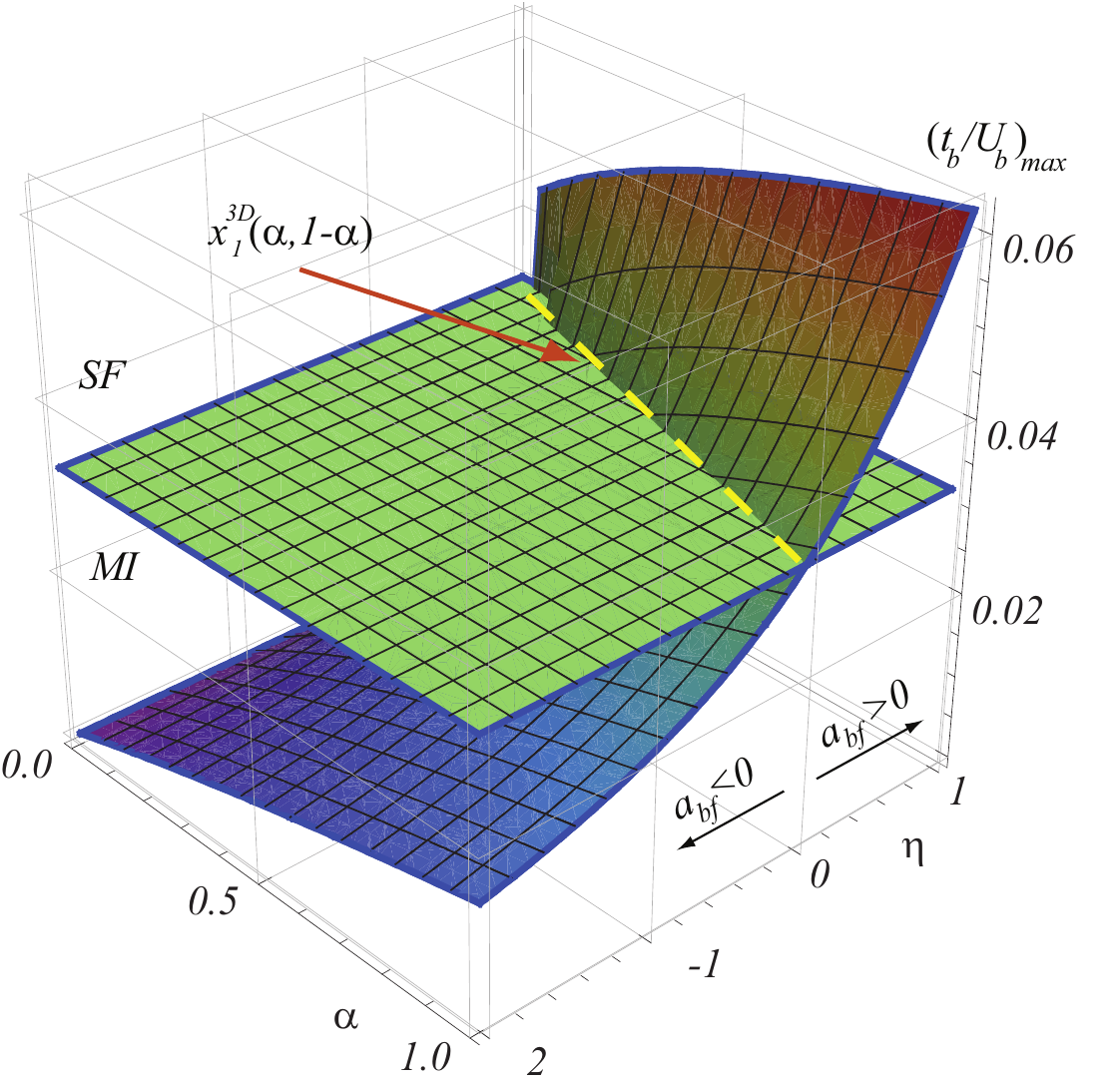}\caption{(Color online) The maximum of the critical value for the parameter
$\left(t_{b}/U_{b}\right)_{\mathrm{max}}$ with the negative real
part of the local dynamic Lindhard function $ $$\chi'\left(\omega\right)<0$
($\omega<\omega_{\mathrm{crit}}$), for cubic ($3D$) lattice in the
space of the dimensionless parameters $\alpha-\eta$ with $n_{\mathrm{B}}=1$.
The flat surface stands for $x_{1}^{3D}\left(1,0\right)$ (see also
Eq. \ref{condition}) in case of the one-component Bose-Hubbard model
with one particle per lattice site. The dashed line stands for the
condition where the system of gaseous mixtures has the same value
$\left(t_{b}/U_{b}\right)_{\mathrm{max}}$ as only bosons confined
in optical lattice. Above the surfaces the superfluid phase takes
place with $\Psi_{\mathrm{B}}\neq0$.}
\label{fig10}
\end{figure}

\begin{figure}
\includegraphics[scale=0.8]{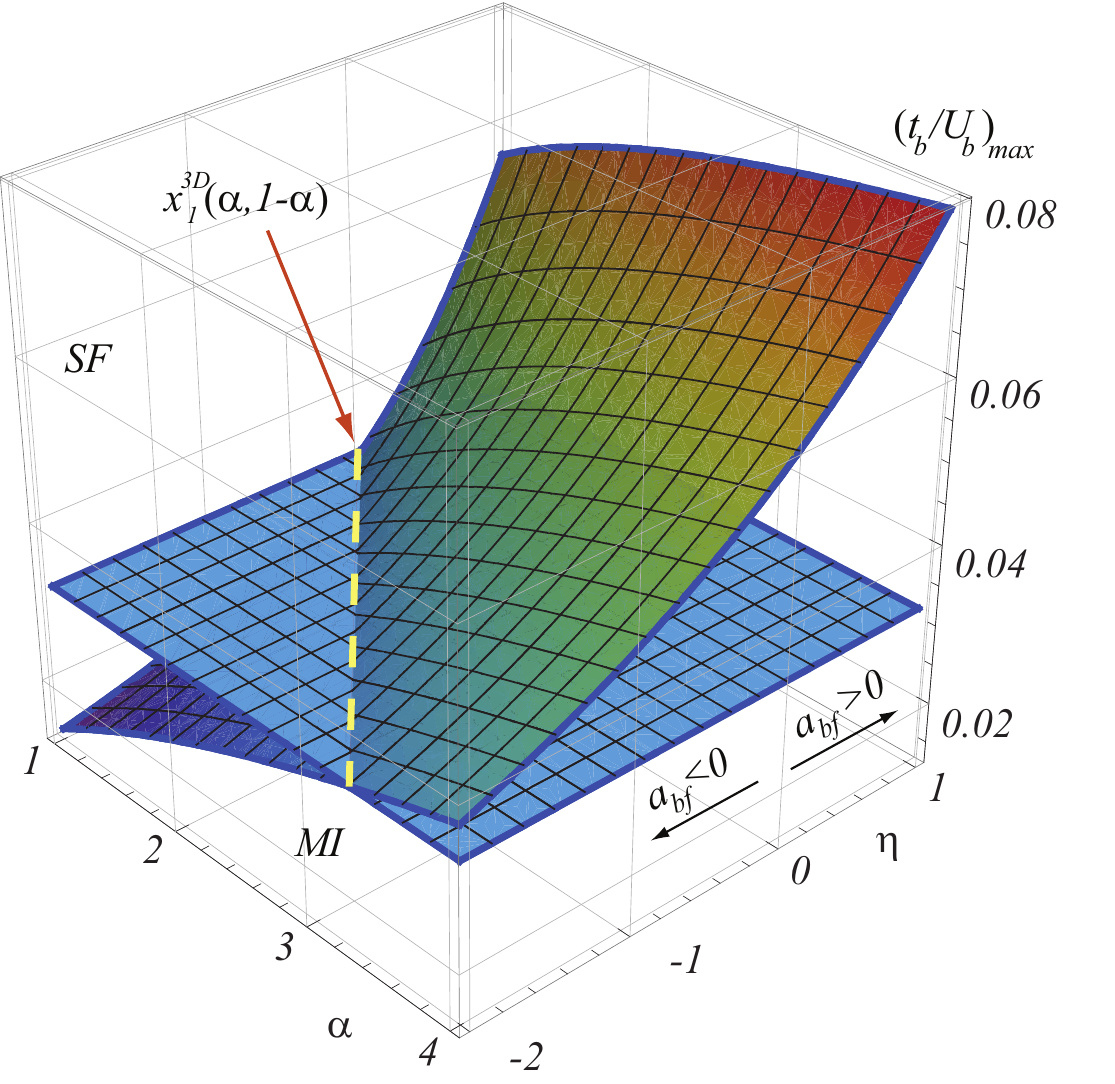}\caption{(Color online) The maximum of the critical value for the parameter
$\left(t_{b}/U_{b}\right)_{\mathrm{max}}$ with the positive value
of the real part of the local dynamic Lindhard function $ $$\chi'\left(\omega\right)>0$
($\omega>\omega_{\mathrm{crit}}$), for cubic ($3D$) lattice in the
space of the dimensionless parameters $\alpha-\eta$ with $n_{\mathrm{B}}=1$.
The flat surface stands for $x_{1}^{3D}\left(1,0\right)$ (see also
Eq. \ref{condition}) in case of the one-component Bose-Hubbard model
with one particle per lattice site. The dashed line stands for the
condition where the system of gaseous mixtures has the same value
$\left(t_{b}/U_{b}\right)_{\mathrm{max}}$ as only bosons confined
in optical lattice. Above the surfaces the superfluid phase takes
place with $\Psi_{\mathrm{B}}\neq0$.}
\label{fig11}
\end{figure}

We want to stress that one have to be careful with the analysis of
the phase diagrams. The summary of our results for square lattice
(we omit qualitatively similar results for cubic geometry) is presented
on Fig (\ref{fig10}) and Fig (\ref{fig11}). For the static and dynamic
limit, but below the critical frequency $\omega<\omega_{\mathrm{crit}}$
the Mott insulator region on the phase diagram broadens only when
the scattering length is positive (the part of the surface above the
plane of the critical value of $x_{1}\left(1,0\right)$ for the one-component
BH model). When we must take into account the difference in the inter-species
masses $m_{b}/m_{f}\neq1$ the sign of the real part of the local
Lindhard response function may become positive and, in consequence,
$\alpha$ parameter takes values above one Fig. (\ref{fig11}) leading
to higher repulsive energy between bosons even if measured scattering
length is negative. In that case for $\alpha>1-\eta$ there is always
a shift for higher values for the parameter $x_{n}\left(\alpha,\eta>1-\alpha\right)$
results in stronger localization of the bosons after adding fermions
to the system. %

\section{Conclusions}

It is well known that the ground state of a system of repulsively
interacting bosons confined in a periodic potential can be either
in a superfluid or in a Mott-insulting state, characterized by integer
boson densities. Because the phase of the order parameter and the
particle number, as conjugate variables, are subject to the uncertainty
principle $\Delta\phi\Delta n\sim\hbar$, so the bosons can either
be in the eigenstate of particle number or phase. The eigenstate of
phase is a superfluid and that of particle number is a localized Mott
insulator. Therefore, the quantum MI-SF phase transition takes place
as the particle density is shifted thus facilitating emergence of
the superfluid from the Mott insulating state. Adding to bosons particles
of different statistics and allowing for the mutual repulsion or attraction
between species strongly affects the equilibrium properties. We presented
a field-theoretic study of the ground-phase diagram in quantum two-
and three-dimensional gaseous Bose-Fermi condensates where mentioned
emulation takes place. We calculated the phase diagram using the quantum
rotor approach that can reproduce the asymmetry in a shift of the
MI to SF transition boundary for positive and negative inter-species
scattering length. Analysis of the local dynamic Lindhard function
revealed the critical value of the frequency for the collective excitations,
where the real part of the response function (and in consequence the
interaction between bosons and fermions) alters sign. The choice of
the parameters of the model led to simple condition for the experimentally
accessible parameters within the phase diagram for Bose-Fermi mixtures
is qualitatively the same as for one-component repulsively interacting
Bose system. We also compared the maximum of the critical value for
$t_{b}/U_{b}$ parameter (as a function of the normalized chemical
potential $\mu_{b}/U_{b}$) at the tip of the $n$th MI lobe for square
and cubic lattice with numerical diagrammatic method and found them
in a good agreement especially for higher, experimentally realizable,
filling factors. The nice feature of presented approach, described
in details above, is that all the expressions and handling are analytic.
It is also worth to notice that provided local approximation can be
very useful in various situations whenever the retardation effects
has to be taken into account and we are not interested in effects
caused by non-locality.

\appendix

\section{Local dynamical approach}

The third term of the trace (Eq. \ref{trace}) can be written after
Fourier transform in form \begin{eqnarray}
\mathrm{Tr}\ln\hat{G}_{f+\mathrm{int}}^{-1} & = & \sum_{\boldsymbol{k}\boldsymbol{k}',\ell,\ell'}\frac{\bar{b}_{\boldsymbol{k}-\boldsymbol{k'}}\left(\omega_{\ell}-\omega_{\ell'}\right)b_{\boldsymbol{k}-\boldsymbol{k'}}\left(\omega_{\ell}-\omega_{\ell'}\right)}{t_{f\boldsymbol{k}}+i\nu_{\ell}}\nonumber \\
 & \times & \frac{\bar{b}_{\boldsymbol{k}-\boldsymbol{k'}}\left(\omega_{\ell'}-\omega_{\ell}\right)b_{\boldsymbol{k}-\boldsymbol{k'}}\left(\omega_{\ell'}-\omega_{\ell}\right)}{t_{f\boldsymbol{k}'}+i\nu_{\ell'}}\nonumber \\
 & = & \sum_{\boldsymbol{q}}\chi\left(\boldsymbol{q},i\tilde{\nu}_{\ell},\mu_{f},\beta\right)\nonumber \\
 & \times & \sum_{\ell}\Lambda_{\boldsymbol{q}}\left(\omega_{\ell}\right)\Lambda_{-\boldsymbol{q}}\left(-\omega_{\ell}\right).\end{eqnarray}
In the above we picked up some special value of the frequency $\tilde{\nu}_{\ell}$.
Now, doing the inverse Fourier transform and using gradient expansion\begin{equation}
b_{j}\left(\tau'\right)=b_{i}\left(\tau\right)+\left(\tau-\tau'\right)\partial_{\tau}b_{j}\left(\tau\right)+\mathcal{O}\left[\left(\tau-\tau'\right)^{2}\right]\end{equation}
we obtain local, in the Matsubara-imaginary time, quadratic form of
the trace in the bosonic variables

\begin{equation}
\mathrm{Tr}\ln\hat{G}_{f+\mathrm{int}}^{-1}\rightarrow\chi'\left(i\tilde{\nu}_{\ell},\mu_{f},\beta\right)\sum_{i}\int_{0}^{\beta}d\tau\left[\bar{b}_{i}\left(\tau\right)b_{i}\left(\tau\right)\right]^{2}.\end{equation}
We performed an expansion were not the $b_{j}\left(\tau'\right)$
degree of freedom itself but rather its gradients of $\partial_{\tau}b_{j}\left(\tau\right)$
are assumed to be small. The explicit formula of the imaginary part
for the dynamical Lindhard function is calculated in the next section.

\section{Local (momentum integrated) Lindhard function}

To stay in the local regime and using an analytic continuation $i\nu_{\ell}\rightarrow$$\omega+i\epsilon$
, where $i\epsilon$ comes from the causality relation of the response
function \begin{equation}
\lim_{\epsilon\rightarrow0^{+}}\frac{1}{\omega\pm i\epsilon}=\mathcal{P}\left(\frac{1}{\omega}\right)\pm i\pi\delta\left(\omega\right),\end{equation}
where symbol $\mathcal{P}$ denotes the Cauchy principal value which
prevents divergence when $\omega=\omega'$, we calculate the explicit
value of the imaginary part that is somewhat easier to obtain than
the real part

\begin{widetext}\begin{eqnarray}
\chi''\left(i\nu_{\ell}\rightarrow\omega+i\epsilon,\mu_{f},\beta\right) & = & \sum_{\mathbf{k},\boldsymbol{q}}\frac{f\left[t_{f\boldsymbol{k}},\mu_{f},\beta\right]-f\left[t_{f\boldsymbol{k}+\boldsymbol{q}},\mu_{f},\beta\right]}{t_{f\boldsymbol{k}}-t_{f\boldsymbol{k}+\boldsymbol{q}}-\omega+i\epsilon}\nonumber \\
 & = & \lim_{\epsilon\rightarrow0^{+}}\frac{1}{\left(2\pi\right)^{4}}\int_{-\pi}^{+\pi}d\boldsymbol{q}d\boldsymbol{k}\frac{f\left[t_{f\boldsymbol{k}},\mu_{f},\beta\right]-f\left[t_{f\boldsymbol{k}+\boldsymbol{q}},\mu_{f},\beta\right]}{t_{f\boldsymbol{k}}-t_{f\boldsymbol{k}+\boldsymbol{q}}-\omega+i\epsilon}\\
 & = & \lim_{\epsilon\rightarrow0^{+}}\int_{-\infty}^{+\infty}dx\int_{-\infty}^{+\infty}dy\frac{f\left(x,\mu_{f},\beta\right)-f\left(y,\mu_{f},\beta\right)}{x-y-\omega+i\epsilon}\rho\left(x\right)\rho\left(y\right)\nonumber \\
 & = & \int_{-\infty}^{+\infty}d\omega'\frac{A\left(\omega',\mu_{f}\right)}{\omega'-\omega}\nonumber \end{eqnarray}
\end{widetext}In the low temperature limit $T\rightarrow0$ the Fermi
distribution becomes $f\left(x\right)=1-\Theta\left(x\right)$ and
we write \begin{eqnarray}
A\left(\omega',\mu_{f}\right) & = & \chi''\left(\omega',\mu_{f}\right)\nonumber \\
 & = & \int_{-\infty}^{+\infty}dx\left[\Theta\left(x-\mu_{f}-\omega'\right)-\Theta\left(x-\mu_{f}\right)\right]\nonumber \\
 & \times & \rho\left(x\right)\rho\left(x-\omega'\right)\end{eqnarray}
that satisfies sum rule $\int_{-\infty}^{+\infty}d\omega A\left(\omega\right)=1$.
Therefore, we can also calculate the real part \begin{eqnarray}
\chi'\left(\omega,\mu_{f}\right) & = & \mathcal{P}\int_{-\infty}^{+\infty}\frac{d\omega'}{\pi}\frac{A\left(\omega',\mu_{f}\right)}{\omega'-\omega}\nonumber \\
 & = & \frac{2}{\pi}\int_{0}^{+\infty}\frac{\omega'\chi''\left(\omega',\mu_{f}\right)}{\omega'^{2}-\omega^{2}}d\omega'.\end{eqnarray}
From these results one finds that $\chi'\left(\omega,\mu_{f}\right)$
is an even function of frequency while $\chi''\left(\omega,\mu_{f}\right)$
is odd.

\section{Fermionic number of particles in the effectively interacting system}

In the non-interacting case the number of fermions on the lattice
can be calculated as follows

\begin{eqnarray}
n_{\mathrm{F}} & = & \frac{1}{N}\sum_{\boldsymbol{k}}\frac{1}{\exp\left[\beta\left(t_{f\boldsymbol{k}}-\mu_{f}\right)\right]+1}\nonumber \\
 & \overset{T\rightarrow0}{=} & \frac{1}{N}\sum_{\boldsymbol{k}}\int_{-\infty}^{+\infty}d\xi\delta\left(\xi-t_{f\boldsymbol{k}}\right)\left[1-\Theta\left(\xi-\mu_{f}\right)\right]\nonumber \\
 & = & 1-\int_{-\infty}^{+\infty}d\xi\rho\left(\xi\right)\Theta\left(\xi-\mu_{f}\right).\end{eqnarray}
with $\rho\left(\xi\right)$ being the density of states for chosen
lattice geometry. Introducing a shift Eq. (\ref{shift}) in the above
we are able to obtain the particular value of the chemical potential
for fermions with effective interaction induced by the coupling with
bosonic species.
\begin{acknowledgments}
We are grateful to N. Teichmann for providing the diagrammatic perturbation
theory data used in Table. We thank R. Micnas for fruitful, stimulating
discussions that allowed to improve some parts of the paper.
\end{acknowledgments}
\appendix

\end{document}